\documentclass[prr,twocolumn,showpacs,superscriptaddress]{revtex4-1}
\usepackage[english]{babel}
\usepackage[dvips]{graphicx}
\usepackage{color}
\usepackage{latexsym}
\usepackage{amsmath}
\usepackage{amsthm}
\usepackage{amsfonts}
\usepackage{amssymb}
\usepackage{mathtools}
\usepackage{cancel}
\usepackage{algorithmic}

\begin{document}

\newcount\layoout

\newcommand{\figeins}{5} 
\newcommand{\figzwei}{3} 
\newcommand{\sizetwo}{0.23}

\pagenumbering{arabic}

\title{Exponential speedup of incoherent tunneling via dissipation}
\author{D.\ Maile}
\affiliation{Fachbereich Physik, Universit{\"a}t Konstanz, D-78457 Konstanz, Germany}
\affiliation{Institut f\"ur Theoretische Physik and Center for Quantum Science, Universit{\"a}t T{\"u}bingen, Auf der Morgenstelle 14, 72076 T{\"u}bingen, Germany}
\author{S. Andergassen}
\affiliation{Institut f\"ur Theoretische Physik and Center for Quantum Science, Universit{\"a}t T{\"u}bingen, Auf der Morgenstelle 14, 72076 T{\"u}bingen, Germany}
\author{W. Belzig}
\affiliation{Fachbereich Physik, Universit{\"a}t Konstanz, D-78457 Konstanz,  Germany}

\author{G. Rastelli}
\affiliation{Fachbereich Physik, Universit{\"a}t Konstanz, D-78457 Konstanz, Germany}
\affiliation{Zukunftskolleg, Universit{\"a}t Konstanz, D-78457, Konstanz, Germany}
\affiliation{INO-CNR BEC Center and Dipartimento di Fisica, Universit{\`a} di Trento, I-38123 Povo, Italy}
\begin{abstract}
We study the escape rate of a particle in a metastable potential in presence of a dissipative bath coupled to the momentum of the particle. 
Using the semiclassical bounce technique, we find that this rate is exponentially enhanced. In particular, the influence of momentum dissipation 
depends on the slope of the barrier that the particle is tunneling through. 
We investigate also the influence of  dissipative baths coupled to the position, and to the momentum of the particle, respectively. 
In this case the rate exhibits a non-monotonic behavior as a function of the dissipative coupling strengths. 
Remarkably, even in presence of position dissipation, momentum dissipation can  
enhance exponentially the escape rate in a large range of the parameter space. 
The influence of the momentum dissipation is also witnessed by the substantial increase of the average energy loss during inelastic (environment-assisted) tunneling. 
\end{abstract}
\date{\today}
\maketitle

%
%
%
%
\section{Introduction} 
In  minimization methods, the research of the absolute minimum becomes a challenging problem when the landscape is characterized by many relative minima and energy barriers of comparable size. 
Classical methods used by classical computers (e.g. Monte Carlo simulations)  generally require exponentially long times.
Quantum adiabatic annealing methods propose an alternative strategy which is based on the idea to elevate the classical system to the quantum domain \cite{Tosatti2006,Stella2005,Lidar20182,Hauke2020,Durkin2019}.
However, such a strategy also poses time constraints. 
An alternative and optimal strategy could be to exploit the quantum tunneling effect as an irreversible process  \cite{Neven2016} 
by avoiding possible quantum coherent oscillations between different minima. 
This can be in principle realized if the quantum system is not closed but is dissipatively coupled to an external bath. 
The influence of dissipation on quantum annealing has been recently studied  \cite{Lidar2015,Hen2017,Lidar2018}.
In open quantum systems, contrary to intuition,  dissipation and dephasing can even enhance
the 
rate of some processes, as in 
quantum computation via engineered dissipation  \cite{Verstraete2009} 
or in transport phenomena assisted by noise in quantum networks and biomolecules \cite{Plenio_2008}
and in photosynthetic biomolecules \cite{Mohseni2008}.
%
%
Environment-assisted quantum transport has been recently studied in a controlled fashion in a spin network formed by trapped ions \cite{Maier2019}.
%

%
%
%
%
\begin{figure}[b!]
	\centering
	\includegraphics[scale=0.433]{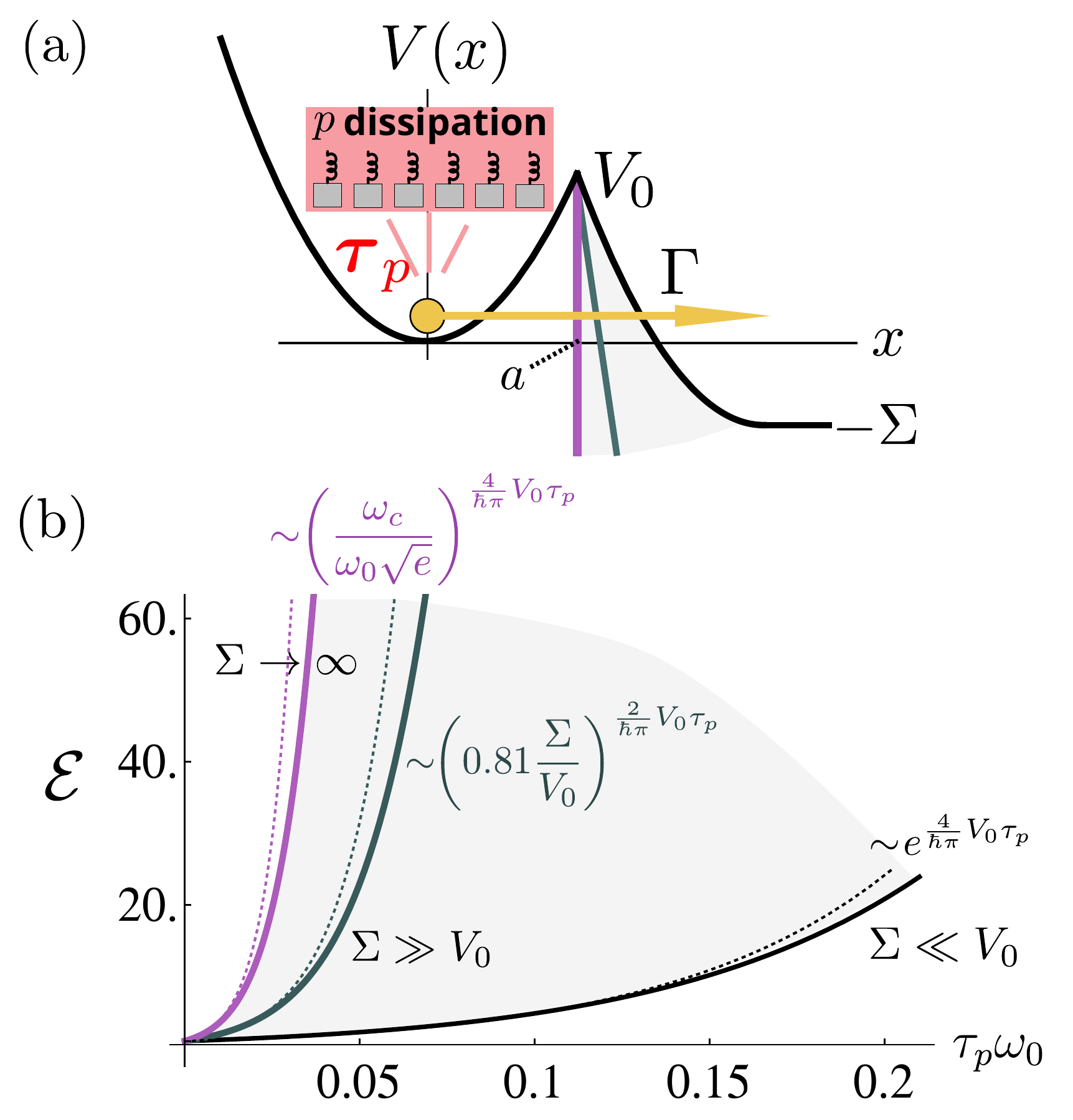}
	\caption{
(a) 
The considered model of a particle in a (semi-double parabolic) potential and
coupled to an external bath via the momentum operator with coupling constant $\tau_p$. 
The slope of the energy barrier on the right can be changed by varying the  
minimum $\Sigma$,  
the different colored lines correspond to different values of $\Sigma$. 
(b) 
Exponential enhancement $\mathcal{E}$ of the escape rate $\Gamma$ as a function of $\tau_p$.
Different lines correspond to different regimes: 
$\Sigma=0.01 V_0$, $\Sigma=10^4 V_0$, $\Sigma=\infty$;  
the dotted lines correspond to analytic expansions (see text).
The frequency of the harmonic well is $\omega_0$ whereas $\omega_c$  is the high frequency cutoff of the bath spectral density.
In all cases the escape rate increases exponentially.
Parameters $V_0/\hbar \omega_0 = 12.5$ and $\omega_c = 8000\, \omega_0$.
}
\label{Fig1}
\end{figure}
%
%
%
%
%
%
%
%
Contrary to the classical case, an isolated quantum particle at zero temperature in a metastable minimum can escape into the unbound region of the potential,  
with a continuous energy spectrum, beyond the energy barrier \cite{Gamow1928,Sethnaimp1982,Callan1977,Coleman1978ae}.  
In order to investigate  
the crossover from the quantum to the classical regime, 
Caldeira and Leggett first considered   
the metastable escape in presence of a dissipative bath coupled to the position \cite{Caldeira1982ann} 
finding that a such coupling leads, indeed,  to a reduction of the escape rate, as intuitively expected.
Subsequent works in the literature analyzed in details this seminal problem \cite{Grabert1984,Grabert1987meta,Grabert19842,Freidkin1986,Risborough1985}.
On the other hand, the problem of the quantum escape from a metastable well in presence of a dissipative bath coupled to the momentum of the particle 
has not been studied so far, although previous studies briefly mentioned a possible enhancement  \cite{Leggett1984prb,Ankerhold2007}.

We here investigate the escape rate 
via quantum tunneling of a   particle trapped in a metastable well and coupled to an external bath via the momentum 
operator, see Fig. \ref{Fig1}.
This kind of dissipative interaction has been discussed in other potentials \cite{Cuccoli2001,Kohler2006ky,Ankerhold2007,Cuccoli2010dr,Kohler2013ie,Rastelli2016ge,Maile2020}.
We here show that  
the coupling to an external bath via the momentum 
operator
increases the escape rate. 
The enhancement is obtained  
even in presence of a second dissipative bath coupled to the position operator of the particle. 
We note that although we analyze an irreversible escape process,  our results are valid also for asymmetric double well potentials 
\cite{Maile2020,Grabert1987} 
as long as  tunneling is completely  incoherent (irreversible) in the regime of strong dissipative coupling with the two baths 
and in the limit of low temperature. 
We also calculate the average energy loss towards the environments during the tunneling process and find a crossover in the dominance of the respective environmental couplings as a function of the slope of the barrier. 
%

\section{Theoretical model}  
In the semiclassical limit $V_0\gg \hbar \omega_0$,  
where $V_0$ is the energy barrier and $\omega_0$ 
the harmonic frequency associated to the relative minimum, 
the escape rate  $\Gamma$ of the particle  through the barrier is of the form 
%
%
%
\begin{align}
\Gamma \, =  \, K \,\, e^{-\frac{1}{\hbar}S_{cl}}\; .
\label{Eq.Gammaescape}
\end{align}
%
%
%
%
%
In the path integral formalism, the  exponent $S_{cl}$ represents the Euclidean action on the minimizing (classical) path $x_{cl}(\tau)$ in the imaginary time  $\tau$ 
and the prefactor $K$ is related to the first order corrections due to fluctuations around this path \cite{Langer1967,Kleinert1995,Weiss2012}. 
In absence of any coupling to an external bath, we denote the  bare escape rate associated to the potential by $\Gamma_0=  K_0 e^{-\frac{1}{\hbar}S_{cl}^{(0)}  }$. 
Generally, the prefactor depends on the dissipative couplings.
While $K$ can be  
enhanced in presence of momentum dissipation affecting the trapped particle, i.e. $K \gg K_0$ (see Appendix \ref{App:C}), the leading 
dependence is due to  
the exponential term.
For this reason  
the present analysis is focused on the ratio between the two exponential terms  
%
%
%
%
%
%
%
\begin{align}
\mathcal{E}
=e^{-\frac{1}{\hbar}\left(S_{cl}^{\phantom{(0)} } - \, S_{cl}^{(0)}  \right)}  \; .
\end{align}
For $\mathcal{E}>1$ the escape rate is increased due to the presence of dissipative couplings, whereas it is reduced for $\mathcal{E}<1$.

Choosing a suitable 
shape of the metastable potential allows us to obtain analytic results also in presence of dissipation. 
For simplicity, 
we model the potential by $V(x)= m \omega_0^2 x^2/2$  for $x<a$  and 
$V(x)=m \omega_0^2 {(x-x_{m})}^2 /2- \Sigma$ for  $a<x<x_{m}$, as shown in Fig.~\ref{Fig1}a. 
From the second minimum the potential remains flat, with $V(x)=-\Sigma$ for $x \geq x_{m}$ which controls the slope on the right side of the barrier.
The matching condition at  
$x=a$ yields
$m \omega_0^2 {(a-x_{m})}^2/2 - \Sigma = V_0$, 
relating $x_{m}$ to $a$.
We also calculated the action 
for a more general potential, showing that the obtained results 
represent the general case, see Appendix \ref{App:E}.
%
%


Integrating out the baths to which the particle is coupled via the momentum and position operators, 
the action in the exponential of Eq.~(\ref{Eq.Gammaescape}) can be split into two parts  $S=S_0+S_{dis}$, with \footnote{The full Hamiltonian is displayed in Appendix \ref{App:A}}
%
%
%
%
%
%
%
\begin{align}
S_0[x(\tau)]
=
\int_{-\frac{\beta}{2}}^{\frac{\beta}{2}}d\tau\left[\frac{m}{2}\dot{x}^{2}(\tau)+V[x(\tau)]\right] 
\label{Eq.metaactionnodis}
\end{align}
%
%
%
%
and the dissipative part as  \cite{Maile2020}
%
%
%
%
%
\begin{align}
S_{dis}[x(\tau)]&
=\frac{1}{2}\iint_{-\frac{\beta}{2}}^{\frac{\beta}{2}}d\tau d\tau'F^{(x)}(\tau-\tau')x(\tau)x(\tau')\nonumber\\&
+\frac{1}{2}\iint_{-\frac{\beta}{2}}^{\frac{\beta}{2}}d\tau d\tau'{F}^{(p)}(\tau-\tau')\dot{x}(\tau)\dot{x}(\tau')\;, 
\label{Eq.dissaction}
\end{align}
%
%
%
%
%
%
%
where the limit $\beta\rightarrow \infty$ has to be performed at the end. 
Assuming Ohmic spectral densities for the two baths,
the two time-dependent functions read
 $F^{(x)}(\tau) = \sum_{l} F^{(x)}_l  e^{i \omega_{l} \tau} / \beta$ and
 ${F}^{(p)}(\tau) = \sum_{l} F^{(p)}_l  e^{i \omega_{l} \tau} / \beta$, 
 with the Matsubara frequency components 
$F^{(x)}_l=\gamma m  |\omega_l| \, f_c( \omega_l ) $
and 
${F}^{(p)}_l= m [ -1 + (1+   \tau_p  |\omega_l|  \, f_c( \omega_l ) )^{-1} ] $ respectively, 
where  $\omega_{l} = 2\pi l /\beta$ (with $l$ integer) \cite{Maile2020} and
$f_c(\omega_l)=  (1+ |\omega_l|  / \omega_c )^{-1}$ is a Drude cutoff function with a frequency cutoff $\omega_c$. We  assume a large frequency cutoff $\omega_c \gg \omega_0$, $\omega_c \gg  \gamma$ and $\omega_c \gg \tau_p \omega_0^2$.
The parameters $\gamma$ and $\tau_p$ are the coupling constants associated to the position and the momentum dissipation, respectively.

%
%
%
%
\begin{figure}[b!]
	\centering
	\includegraphics[scale=0.3]{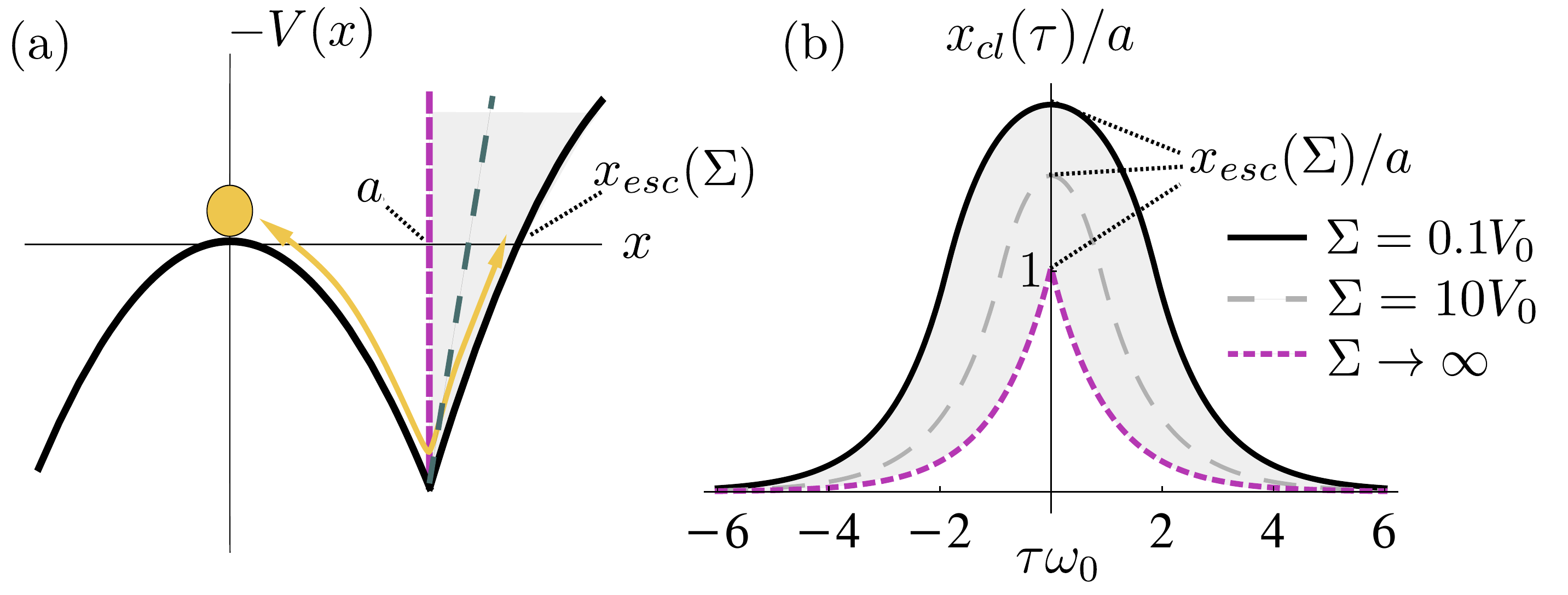}
	\caption{ 
	(a) Example of the inverted potential $-V(x)$ for the motion of the minimizing (classical) path $x_{cl}(\tau)$. 
	(b) Different paths  $x_{cl}(\tau)$ for different values of $\Sigma$ in the non dissipative case $\gamma=\tau_p=0$.
	 The path shrinks with increasing $\Sigma$. For $\Sigma = \infty$ the particle is instantly reflected at the turning point $x_{esc}$. 
	 }
	\label{Fig2}
\end{figure}
%
%
%

In the semiclassical path integral method for the quantum decay, 
the minimizing  path $x_{cl}(\tau)$ of  the action $S$ is the solution of the classical equation 
of motion with the inverted potential $-V(x)$.
At zero temperature ($\beta\rightarrow \infty$) a non-trivial solution $x_{cl}(\tau) \neq 0$ is called bounce path: 
The particle starts at the minimum of the well at $x(\tau\!=\!-\beta/2)|_{\beta \rightarrow \infty} =0$, reaches the turning point $x(\tau\!=\!0) = x_{esc}$, 
and then returns to $x(\tau\!= \!\beta/2)|_{\beta \rightarrow \infty} =0$.
Examples of $x_{cl}(\tau)$ for the bare potential $\gamma=\tau_p=0$  
are reported in Fig.~\ref{Fig2}, 
for the ones in presence of dissipation we refer to the Appendix \ref{App:B}.
The imaginary time spent in the region $a<x(\tau)<x_{esc}$ is called bounce time $\xi_B$, which turns out to depend strongly on the slope of the potential and ultimately vanishes in the limit of sharp potential. 
This characteristic time scale is determined by the integral equation \footnote{We  find the bounce path by inserting the ansatz 
$
x_{cl}(\tau)=\frac{1}{\beta }\sum_{l} x_{l} e^{i\omega_{l}\tau} / \beta 
$
into the action $S$ and minimizing it with respect to $x_l$.} 
%
%
%
%
%
\begin{align}
\frac{1}{\omega_0^2}&\frac{1}{\sqrt{1+\Sigma/V_0}+1} = \label{Eq.timeintegraldis0} \\ 
&
 \frac{1}{\pi}\int_{0}^{\infty}d\omega\frac{\sin(\omega\xi_{B})}{\omega\left(\frac{\omega^{2}}{1+\tau_{p}\omega f_c(\omega)}+\omega_{0}^{2}+\gamma\omega f_{c}(\omega)\right)}\ 
\;. \nonumber
\end{align}
%
%
%
%
%
%
The analytic solution for the action $S_{cl} $ is given by 
%
%
%
%
%
\begin{align}
S_{cl} 
& =-\frac{2\omega_{0}^{2}V_0\left(\sqrt{1+\Sigma/V_0}+1\right)^{2}}{\pi} \nonumber \\ &\times \int_{0}^{\infty}d\omega\frac{1-\cos\left(\omega\xi_B\right)}{\omega^{2}\left(\frac{\omega^{2}}{1+\tau_{p}\omega f_c(\omega)}+\omega_{0}^{2}+\gamma\omega f_{c}(\omega)\right)} \nonumber \\ &+2V_0\left(1+\sqrt{1+\Sigma/V_0}\right)\xi_{B}\;.
\label{Eq.Actiongeneral0}
\end{align}
%
%
%
%
%
We note that the dependence on the bounce time $\xi_B$ implies a change of the action when the slope of the right side of the potential barrier  
is varied.

%
%
%
%
\begin{figure*}[t]
	\centering
	\includegraphics[scale=0.23]{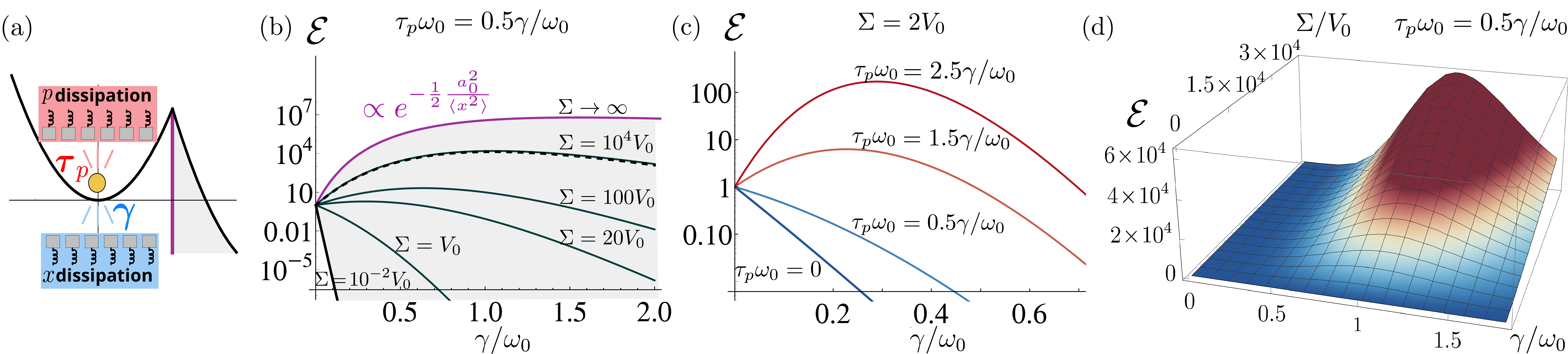}
	\caption{
	 (a) Particle in a metastable well in presence of both dissipative couplings. 
	 (b) Logarithmic plot of $\mathcal{E}$ as a function of $\gamma$,
	 at fixed ratio $\tau_p\omega_0^2/\gamma=0.5$ and for different values of $\Sigma$. 
	 The gray area and the violet and black lines correspond to the shape in the panel (a).
	 The dashed line is the analytical result shown in the Appendix \ref{App:B}. 
	 (c) Logarithmic plot of $\mathcal{E}$ as a function of $\gamma$, 
	for  $\Sigma = 2 V_0$ and different values of the ratio  $\tau_p\omega_0^2/\gamma$. 
	 (d) $\mathcal{E}$ as a function of $\Sigma$ and $\gamma$, at fixed ratio  $\tau_p\omega_0^2/\gamma=0.5$.
	For (b), (c) and (d) we used $V_0/\hbar\omega_0 = 12.5$.}
	\label{Fig3}
\end{figure*}
%
%
%
%
%
%
%

\section{Results} 
Our main results are summarized in Figs.~\ref{Fig1} and \ref{Fig3}.
In Fig.~\ref{Fig1}, we show the results for pure momentum dissipation of coupling strength $\tau_p$, 
for different slopes of the right side of the barrier.
In presence of pure momentum dissipation, the action decreases as a function of the coupling parameter $\tau_p$  
leading to the enhancement of the escape rate observed in Fig.~\ref{Fig1}b.
In contrast, for pure position dissipation in a metastable well one obtains an exponential 
suppression of the escape rate \cite{Caldeira1982ann,Grabert1987meta,Grabert1984,Freidkin1986,Risborough1985}. 

A simple picture for the observed exponential speedup of the escape rate can be given in the limit of
infinite slope (see Fig.~\ref{Fig1}a),  which is obtained by taking the limit  $\Sigma\rightarrow \infty$ \cite{Grabert1987meta}.
%
%
As a consequence, the particle is instantly reflected at $x_{esc} = a$ leading to a vanishing  bounce time $\xi_B  \rightarrow 0$ (see Fig.~\ref{Fig2}b).
In this case, $S_{cl}/\hbar \equiv a^2 / (2 \left<  x^2 \right>)  $, with $a$ the position of the barrier maximum 
and $ \left<  x^2 \right>  $ the harmonic quantum fluctuations of the particle in the well.
As the momentum dissipation {\sl increases} the quantum fluctuations $ \left<  x^2 \right> $
\cite{Cuccoli2001,Kohler2006ky,Rastelli2016ge}, the escape rate is enhanced as a consequence.
In the weak coupling limit $\omega_0 \tau_p \ll1$, the expansion of $ \left<  x^2 \right>$ in the action $S_{cl}$ yields
%
%
%
%
%
\begin{equation}
\mathcal{E}_{\gamma=0}^{(\Sigma=\infty)}
\!
\approx
\!  
 \left(  \frac{\omega_c}{\omega_0\sqrt{e}} \right)^{ \frac{4 }{\pi \hbar} \, V_0 \tau_p }
 \,\,\, \text{ } \,\,  
  \,\,  \omega_0\tau_p\ll 1 , \,\,\, \omega_c \xi_B \ll1 . \label{Eq.approxuncon0} 
\end{equation} 
The escape rate $\mathcal{E}$ is limited by the cutoff frequency $\omega_c$ as $ \left<  x^2 \right>$  diverges in the limit $\omega_c \rightarrow \infty$ \cite{Rastelli2016ge}. 
This simple form provides a good approximation also for a sharp barrier with a finite slope (and small $\xi_B$) as long as $\omega_c \xi_B \ll1$.

For a finite slope of the barrier, the classical action is not anymore simply related to the harmonic quantum fluctuations $S_{cl} / \hbar \neq a^2 / (2 \left<  x^2 \right>) $.
However, the exponential enhancement still persists in this more realistic situation shown in Fig.~\ref{Fig1}b and in particular is not affected by the presence of conventional (position) dissipation, as will be discussed in the following. 
The exponential enhancement of the rate in presence of momentum dissipation strongly depends on the steepness of the barrier.
As shown in Fig.~\ref{Fig1}b, the effect is maximal for a sharp barrier $(\Sigma=\infty)$, which represents an upper theoretical bound, and then becomes less pronounced.
A qualitative understanding is provided by the following physical picture: 
The presence of the momentum dissipation gives rise to an anomalous friction force in the equation for the minimal path  $x_{cl}$.
Such a force is proportional 
to the acceleration through a memory  friction function  (i.e. nonlocal in time), see Appendix \ref{App:B}.
As the acceleration is strongly controlled by the conservative force $\sim dV(x)/dx$, 
the anomalous friction force becomes important at the turning point $x_{esc}$, 
viz. when $dx^2(\tau)/d^2\tau$ becomes large, see example of Fig.~\ref{Fig2}. 
Notice that the turning point $x_{esc}$ also depends on the dissipative couplings and does not coincide with the zero of the potential $V(x_0)=V(0)=0$ as in the case of 
no dissipation (see Appendix \ref{App:B}).
Finally, we remark that even in the limit ${\left. dV(x)/dx \right|}_{x=x_{esc}} = -\infty$, the results for the action remain finite as the friction is nonlocal in time.

For finite value of $\Sigma$, an intermediate regime is identified by 
a reduced slope of the potential for $\Sigma \gg V_0$ and an increased bounce time $ \omega_c  \xi_B \gg 1$
 (but still  $\omega_0 \xi_B \ll  1$).
In this case, the action is not limited anymore by the cutoff frequency $\omega_c$. 
The escape rate still depends exponentially on the dissipative momentum interaction, but with an explicit dependence on $\Sigma$
%
%
%
%
%
\begin{align}
\mathcal{E}_{\gamma=0}^{(\Sigma \gg V_0)}
\approx  \left(k_1\frac{{\Sigma}}{V_0}\right)^{ \frac{2 }{\pi \hbar} \, V_0 \tau_p }
\,\,\, \text{ } \,\,  
  \,\,  \omega_0\tau_p\ll 1 , \,\,\, \omega_c \xi_B \gg 1 , 
\; \label{Eq.approxuncon}
\end{align}
%
%
%
%
%
%
where $k_1=e^{2(1-C)}/4\approx 0.81$  with $C$ the Euler constant. 
Eq.~(\ref{Eq.approxuncon}) corresponds to the (gray) dotted line in Fig.~\ref{Fig1}b.

A further interesting regime corresponds to $\Sigma\ll V_0$, for a finite slope and bounce time 
\footnote{
In the limit $\Sigma\ll V_0$, the action is closely related to the problem of the double well studied in \cite{Maile2020} 
because the potentials are equal in the region $x<x_m$.}.
In the limit $\Sigma \ll V_0$ the problem becomes equivalent to the solution of a slightly asymmetric double 
well in the incoherent overdamped limit, discussed in \cite{Grabert1987} for pure position dissipation.
In a such limit the exponential enhancement is described by
%
%
%
%
%
\begin{align}
\mathcal{E}_{\gamma=0}^{(\Sigma \ll V_0)}
\approx  e^{ \frac{4 }{\pi \hbar } \, V_0 \tau_p }
\,\,\, \text{ } \,\,  
  \,\,  \omega_0\tau_p\ll 1 , \,\,\, \omega_0 \xi_B \gg 1 . 
\; \label{Eq.approxuncon2}
\end{align}
%
%
%
%
%
%
We note that the different regimes of
$\Sigma$ described by Eqs.~(\ref{Eq.approxuncon0})-(\ref{Eq.approxuncon2}) present characteristic base functions, whereas the exponent controlling the effect of momentum dissipation is the same.

The analytical solutions  for $\mathcal{E}$ presented so far are restricted  to  pure momentum dissipation  $(\gamma=0)$.
The respective curves are displayed in Fig.~\ref{Fig1} and agree well with the numerical results of Eqs.~(\ref{Eq.timeintegraldis0}) and (\ref{Eq.Actiongeneral0}) in the respective regimes.
Further analytical expressions in presence of both dissipative couplings are reported in the Appendix \ref{App:B} 
in various regimes  of  $\omega_0 \xi_B$,  $\omega_c \xi_B$ and $\Sigma / V_0$ and they also agree well with the numerical results 
(an example is given in Fig.~\ref{Fig3}b). 
The main result analyzed for pure momentum dissipation also holds in presence of both baths when position dissipation dominates, as shown in Fig.~\ref{Fig3}.
In general, the presence of both dissipative couplings (momentum and position)  
leads however to a non-monotonic  
behavior as a function of $\gamma/\omega_0$ (or $\tau_{p}\omega_0$) clearly visible in $\mathcal{E}$ shown in Fig.~\ref{Fig3}b-d for a fixed ratio $\tau_{p}\omega^2_0/\gamma$.  
Again, a simple physical picture is obtained in the limit $\Sigma=\infty$ in which  $\mathcal{E}$  is determined uniquely by the harmonic quantum fluctuations $ \left<  x^2 \right> $ 
of the harmonic well.
Such harmonic quantum fluctuations  exhibit a non-monotonic behavior as a function of the   
coupling strengths \cite{Rastelli2016ge}. 
This is reflected in the results for $\mathcal{E}$ shown in Fig. \ref{Fig3}b for a 
fixed ratio $\tau_p\omega_0^2 / \gamma$
and different values of $\Sigma$.
We note that, in the regime of sharp potential $\Sigma \gg V_0$, the effect of momentum dissipation dominates over the range of $\gamma/\omega_0$ shown in the figure.
This is due to the strong dependence of the momentum dissipation on the slope as explained previously.
In Fig.~\ref{Fig3}c we show $\mathcal{E}$  for a given value of $\Sigma=2V_0$ and for different values of the ratio $\tau_p\omega_0^2/\gamma$.
We observe that the exponential speed-up still persists in a wide range of  $\gamma/\omega_0$  which depends on $\tau_p\omega_0^2/\gamma$.

\begin{figure}[b]
	\centering
	\includegraphics[scale=0.40]{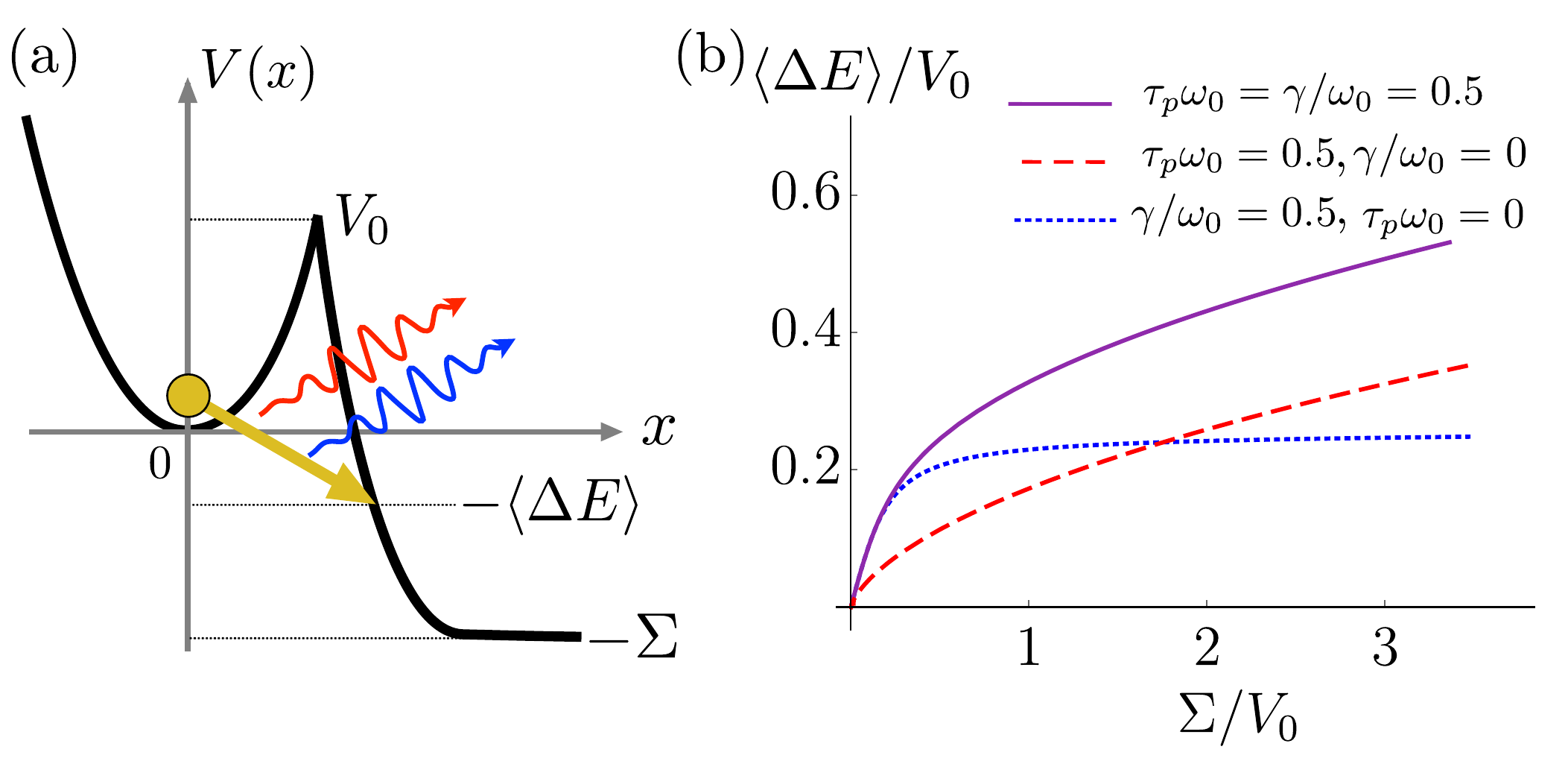}
	\caption{
	 (a) Average energy loss during the tunneling process $\langle \Delta E \rangle=V(0)- V(x_{esc})$ for a particle coupled 
	to two baths via the momentum and the position operators.  
	(b) The quantity $\langle \Delta E \rangle$ as a function of  $\Sigma$ for different dissipative cases. 
	The solid (purple) line is for both dissipative couplings whereas the dashed (red) line and the dotted (blue) line are in presence of a single bath with coupling through the momentum 
	and the position, respectively. 
		}
	\label{Fig4}
\end{figure}

Finally, we also calculate the average energy loss during the tunneling process. 
Such average energy induced by the dissipative couplings is defined as $\langle \Delta E \rangle=V(0)-V(x_{esc})$ 
\cite{Grabert1987,Weiss1984}.
In absence of dissipative interaction, the returning point $x_{esc}$ of the bounce path is simply given by the condition  $V(x_{esc})=V(0)=0$, see Fig.~\ref{Fig2}a.
However, by increasing the dissipative coupling, $x_{esc}$ shifts to larger values  (see Appendices \ref{App:B} and \ref{App:D}), 
indicating that the particle escapes the barrier, on average, at an energy $V(x_{esc})<V(0)$, see Fig.~\ref{Fig4}a. 
In Fig.~\ref{Fig4}b we plot $\langle \Delta E \rangle$ as a function of $\Sigma$ for different dissipative cases. 
Considering pure position dissipation we find that the loss saturates already for moderate $\Sigma \sim V_0$ becoming independent of the barrier.
%
%
%
%
%
%
%
%
%
%
For small $\Sigma$ the energy loss into a single momentum dissipative bath is smaller than in the presence of pure position dissipation. 
However, the latter energy loss increases by increasing $\Sigma$ and eventually becomes larger as in the position dissipative counterpart.
This is again a consequence of the explained increasing influence of momentum dissipation as a function of $\Sigma$.
By further increasing $\Sigma$, $\langle \Delta E\rangle$  eventually saturates to a value determined by the cutoff frequency $\omega_c$ (see Appendix \ref{App:D} for the details).
In presence of both dissipative couplings and $\Sigma\ll V_0$ the loss coincides with the value of pure position dissipation as can be seen from Fig.~\ref{Fig4}b.
For larger values of $\Sigma$, the energy losses simply add up. 
From the crossing point in Fig.~\ref{Fig4}b we can determine the values in parameter space for which both dissipative baths become equally important.
Further, the average energy loss as a function of the dissipative coupling saturates to the same asymptotic value in the overdamped limit $\gamma/\omega_0 \gg 1$ and $\tau_p\omega_0 \gg 1$ 
as we display in  Appendix \ref{App:D}.  Note that the asymptotic values in Fig.~\ref{Fig4}b satisfy $\langle \Delta E\rangle_a < \Sigma$
since the expected maximum energy loss is $V(0)-V(x_m)=\Sigma$ \cite{Weiss1984}. 


%
%
%
%

\section{Conclusions} 
Momentum dissipation leads to an exponential enhancement of the escape rate of a particle in a metastable potential.
In presence of position dissipation, we find a non-monotonic behavior as a function of the dissipative coupling strengths. 
Depending on the barrier, momentum or position dissipation can be dominant. 
For a sharper barrier, the region of momentum dissipation induced enhancement increases. 
The particle’s average energy loss during the tunneling process shows a strong dependence on the interplay between momentum dissipation and the slope of the potential. 
We verified that the discussed physical implications do not depend on the specific form of the potential and the qualitative behavior corresponds to the general case (as we show in Appendix E)

To summarize, we propose a method that rapidly releases the system from a relative (metastable) minimum exploiting quantum tunneling as a pure, irreversible and inelastic process, assisted by the environment.
Our theoretical findings can be directly tested in superconducting quantum circuits \cite{Pop2018,Manucharyan2019,Rastelli2015} in which dissipative position and momentum interaction translate to dissipative phase or charge couplings. 
In particular,  momentum/charge dissipation can be readily implemented simply using capacitances and resistances \cite{Maile2018}.  Further, our results  are important for quantum numerical minimization methods in which 
the escape rate from a relative minimum plays a key role in setting the computational time scale  \cite{Neven2016,Smelyanskiy2016}. \\


\acknowledgments
We acknowledge
Financial support from the MWK-RiSC program.
This research was also partially supported by the German Excellence
Initiative through the Zukunftskolleg and by the Deutsche
Forschungsgemeinschaft (DFG) through the SFB 767 and Project-ID 425217212 – SFB 1432.  
G.R. thanks P. Hauke  for useful discussions and relevant comments.

\appendix
\begin{widetext}

\section{General formula for the decay rate in the semiclassical approximation}\label{App:A}
In this section, we recall the theoretical method for calculating the amplitude (or matrix element) $Z_0$ which can be related to the escape rate of a particle placed in the metastable well, in the zero temperature limit.
We generalize this approach in presence of position and momentum dissipation.

\subsection{The Hamiltonian of the system}
The system discussed in the main text is described the Caldeira-Leggett Hamiltonian with two dissipative harmonic oscillator baths 
\begin{align}
    \mathcal{H}=\mathcal{H}_{sys}+\mathcal{H}^{(x)}_B+\mathcal{H}^{(p)}_B.
\end{align}
Here, $\mathcal{H}_{sys}=p^2/2m+V(x)$ is the system Hamiltonian with momentum $p$ and position $x$, $\mathcal{H}^{(x)}_B$ is the bath Hamiltonian including the bilinear coupling to the position, reading
\begin{align}
\mathcal{H}^{(x)}_{B}=\frac{1}{2}\sum_i\left[\frac{P^2_i}{M_i}+M_i\omega_i^2\left(X_i-\frac{\lambda_i}{M_i\omega_i^2}x\right)^2\right]\label{Eq.CLx},
\end{align}
where $P_i,X_i,M_i$ and $\omega_i$ are the momentum, the position, the mass and the frequency of one harmonic oscillator in the bath, respectively. Further, $\lambda_i$ denotes the linear coupling to $x$, and the counter term cancelling the renormalization of the potential is included.  
$\mathcal{H}^{(p)}_B$ yields a bilinearly coupling to the momentum, reading
\begin{align}
\mathcal{H}^{(p)}_{B}=\frac{1}{2}\sum_j\left[\frac{\left(P_j-\mu_jp\right)^2}{M_j}+M_j\omega_j^2X^2_j\right]\label{Eq.CLp},
\end{align}
where the subscripts $j$ denote the coordinates, the mass and the frequency of the bath coupled to the momentum and $\mu_j$ is the linear coupling strength. Integrating out the bath in the path integral language leads to the action in Eqs.~(3) and (4) of the main text. 
%
%
\subsection{The semiclassical method}
Because the potential is metastable, $Z_0$ has an imaginary part $\Gamma$ which corresponds to the  escape rate.
One can calculate $Z_0$ via the imaginary time path integral method, specifically using the
instanton-bounce method [\onlinecite{Coleman1978ae,Grabert19842,Kleinert1995,Weiss2012,Sethnaimp1982}].
The starting point of this theoretical approach is the amplitude in the imaginary time
\begin{equation}
Z_{0} =\langle x_0 |e^{-\frac{\beta}{\hbar}\mathcal{H}}| x_0 \rangle=\oint D[x(\tau)]e^{-\frac{1}{\hbar}S[x(\tau)]} \; ,
\end{equation}
where $S[x(\tau)]=S_{0}+S_{dis}$ is the action of the open quantum system  given in the main text. In particular  $S_{dis}$ is the dissipative action and  $x(\tau)$ is a generic periodic path with $x_0 = x(\beta/2)=x(-\beta/2)$. 
In the limit $\beta\rightarrow\infty$, one can set $x_0=0$ such that $Z_0$ is proportional to the density of probability (in the imaginary time) to find the particle at the origin, 
which corresponds to the minimum of the metastable well.

One can calculate $Z_0$ in the semiclassical approximation by finding the so-called
classical path $x_{cl}(\tau)$ that minimizes the action and using the expansion $x(\tau)=x_{cl}(\tau)+\delta x(\tau)$
leading to $S[x(\tau)]=S_{cl}[x_{cl}(\tau)]+S_{\delta}[\delta x(\tau)]$. 
This path, in the zero temperature limit $\beta \rightarrow \infty$, 
starts and ends in the minimum of the well, $x(\pm\beta/2)|_{\beta\rightarrow \infty}=0$.
Beyond the trivial solution $x^{(0)}_{cl}(\tau)=0$, there exists the so-called bounce path $x_{cl}^{(1)}(\tau)$  in which the particle moves from the minimum $x=0$, 
gets reflected at the returning point $x_{esc}$ and comes back to its origin, see Fig.~\ref{Fig1-supp}.
The matrix element of a single bounce path $x_{cl}^{(1)}$ can be written in the semiclassical limit as
\begin{equation}
z_0^{(1)}
=
e^{-\frac{1}{\hbar}S_{cl}[x_{cl}^{(1)}(\tau)]}\oint D[\delta x(\tau)]e^{-\frac{1}{\hbar}S_{\delta}^{(1)}[\delta x(\tau)]} \;,
\label{Eq.bouncez1}
\end{equation}
with the fluctuations around the classical path satisfying $\delta x(-\beta/2)=\delta x(\beta/2)=0$ and $S_{\delta}^{(1)}$ being the expansion of the action 
over the single bounce path  $x_{cl}^{(1)}$
\begin{align}
S^{(1)}_{\delta}[\delta x(\tau)]
&=
\int_{-\frac{\beta}{2}}^{\frac{\beta}{2}}d\tau\left(\frac{m}{2}\delta\dot{x}^{2}(\tau)+\frac{1}{2}\frac{d^{2}V(x(\tau))}{d^{2}x}\left|_{x^{(1)}_{cl}} \right.\delta x^{2}(\tau)\!\right) +S_{dis}[\delta x(\tau)]\;.
\label{Eq.differential}
\end{align}
One can express the generic fluctuations path as $\delta x(\tau)=\sum_{q=0}^{\infty}c_{q}y_{q}(\tau)$ in which we use as basis the eigenfunctions 
of the following eigenvalue equation
\begin{align}
&\left[-m\frac{d^{2}}{d\tau^{2}}+\mathcal{V}[x_{cl}^{(1)}]\right]y_{q}(\tau)+\int_{-\frac{\beta}{2}}^{\frac{\beta}{2}}d\tau'F^{(x)}(\tau-\tau')y_{q}(\tau')-\int_{-\frac{\beta}{2}}^{\frac{\beta}{2}}d\tau'F^{(p)}(\tau-\tau')\frac{d^{2}}{d\tau'^{2}}y_{q}(\tau')
=\lambda_{q}^{(B)}y_{ q }(\tau)
\;,
\label{eq:schr}
\end{align}
where we set $\mathcal{V}[x_{cl}^{(1)}]=\frac{d^2V[x(\tau)]}{dx^2}\left|_{x_{cl}^{(1)}}\right.$  and $\lambda_{q}^{(B)}$ are the eigenvalues. 
Using the eigenvectors decomposition, we can write
\begin{equation}
S^{(1)}_{\delta} =\frac{1}{2}\sum_{q=0}^{\infty}\lambda_{q}^{(B)}c_{q}^{2}
\;.
\end{equation}
A priori, the full matrix element is the sum of all possible $n$-bounce paths
\begin{equation}
Z_0=\sum_{n=0}^\infty z_0^{(n)}\;.
\label{eq:sum}
\end{equation}
As discussed below, we will use the so-called dilute gas approximation for the bounces such that the quantities $ z_0^{(n)}$ for $n \geq 2$ can be obtained from the knowledge of $ z_0^{(1)}$.
\begin{figure}[t!]
	\centering
	\includegraphics[width=0.8\linewidth]{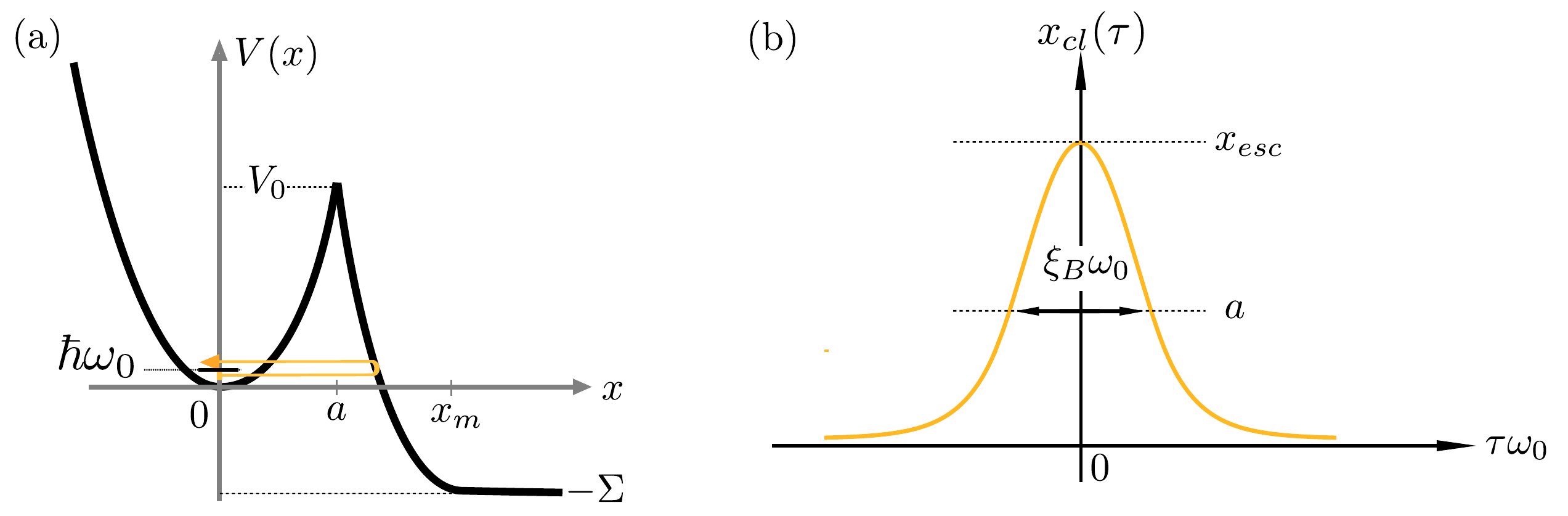}
	\caption{(a) Metastable well potential discussed in the present article. 
	(b) Example of a single bounce path which minimizes the action without dissipation.
	The bounce time $\xi_B$ denotes the imaginary time interval in which 
	 the path is in the region $x > a$.}
	\label{Fig1-supp}
\end{figure}
\mbox{}\\

%
%
\subsection{Discussion on the zero and on the negative eigenvalues}

The spectrum $\{\lambda^{(B)}_q\}$ contains a zero eigenvalue
$\lambda_0^{(B)}=0$ with the eigenfunction $y_0(\tau)=A_{\tau_0}\dot{x}^{(1)}_{cl}(\tau)$ because of the translational invariance of the bounce on the whole $\tau$ axis.  
$A_{\tau_0}$ is a constant and $\dot{x}_{cl}(\tau)=d{x}_{cl}(\tau)/d\tau$.  
There exists also a negative eigenvalue $\lambda_1^{(B)}$
due to the fluctuations of the bounce time $\xi$ 
which corresponds to the imaginary time in which the path is in the region $a < x < x_{esc}$.
From a mathematical point of view, one can consider Eq.~(\ref{eq:schr}) as the  Schr\"{o}dinger differential equation
in which the $\tau$ axis is the space and $y_q(\tau)$ the wavefunction. Then the negative eigenvalue $\lambda_1^{(B)}$  can be seen as the bound energy of the localized  
ground state [\onlinecite{Kleinert1995}].
This leads to an imaginary part of the amplitude $z_0^{(1)}$ as we show below. 

We use the following change of variable by expressing the generic path (in the semiclassical limit) as
\begin{align}
x(\tau )&=x^{(1)}_{cl}(\tau) +\sum_{q=0}^{\infty } c_q y_q(\tau) 
\,\, \longrightarrow \,\, 
x(\tau,\xi )=x^{(1)}_{cl}(\tau-\tau_0, \xi)+\sum_{q=2}^{\infty } \tilde{c}_q y_q(\tau-\tau_0, \xi) 
\;,
\label{Eq.Fullpath}
\end{align}
in which we consider $\tau_0$, the center of the bounce, and $\xi$, the bounce time, as the new variables instead of $c_0$ and $c_1$.
The Jacobian $|dc_0/d\tau_0|$  is obtained via the overlap of $x(\tau,\xi )$ with $y_0(\tau)=A_{\tau_0}\dot{x}^{(1)}_{cl}(\tau)$  yielding
\begin{align}
c_0(\tau_0)=&\int_{-\frac{\beta}{2}}^{\frac{\beta}{2}}d\tau x(\tau,\xi )y_0(\tau,\xi)
=
\int_{-\frac{\beta}{2}}^{\frac{\beta}{2}}d\tau[x_{cl}^{(1)}(\tau-\tau_{0},\xi)A_{\tau_{0}}\dot{x}_{cl}^{(1)}(\tau,\xi)+\sum_{q=2}^{\infty}
\tilde{c}_{q}y_{q}(\tau-\tau_{0},\xi)y_0(\tau)]
\;.
\label{Eq.co1}
\end{align}
Because of the translational invariance we can expand
\begin{align}
x_{cl}^{(1)}(\tau-\tau_{0},\xi)&\approx x_{cl}^{(1)}(\tau,\xi)+\dot{x}_{cl}^{(1)}(\tau,\xi)\tau_{0}
\end{align}
and analogously $y_q(\tau-\tau_{0},\xi) \approx y_q(\tau,\xi)+\dot{y}_q(\tau,\xi)\tau_{0}$.
Inserting these expressions into (\ref{Eq.co1}) we find
\begin{align}
c_0(\tau_0)=& 
\int_{-\frac{\beta}{2}}^{\frac{\beta}{2}}d\tau\! \left[A_{\tau_{0}}\left(x^{(1)}_{cl}(\tau,\xi)\dot{x}^{(1)}_{cl}(\tau,\xi)
+\left(\dot{x}^{(1)}_{cl}(\tau,\xi)\right)^{2}\right)+\sum_{q=2}^{\infty} \tilde{c}_{q}\dot{y}_{q}(\tau,\xi)y_{0}(\tau,\xi)\right]\tau_{0}
\end{align}
and hence
\begin{align}
\left|\frac{dc_{0}(\tau_{0})}{d\tau_{0}}\right|
&=\frac{1}{A_{\tau_{0}}}+A_{\tau_{0}}\int_{-\frac{\beta}{2}}^{\frac{\beta}{2}}d\tau\sum_{q=2}^{\infty} \tilde{c}_{q}\dot{y}_{q}(\tau,\xi)\dot{x}^{(1)}_{cl}(\tau,\xi)
\;,\label{Eq.Jac1}
\end{align}
where we used that 
\begin{equation}
A_{\tau_{0}}=\left(\sqrt{\int_{-\beta/2}^{{\beta/2}}d\tau\left(\dot{x}^{(1)}_{cl}(\tau)\right)^{2}}\right)^{-1}\;,\label{Eq.Atau0}
\end{equation}
which follows from the normalization condition $\int_{-\beta/2}^{{\beta/2}}y^2_0(\tau)d\tau=1$.

%
%
\subsection{Transformation of the fluctuations path integral for  a single bounce}
A priori, we must deal with the integration of the path integral for the fluctuations.
We find for the transformation
\begin{align}
\oint_{\delta x\left(\frac{\pm\beta}{2}\right)=0}  D[\delta x(\tau)] 
\,\, \longrightarrow \,\,\,  
 \mathcal{N} \prod_{q=0}^{\infty} \int_{-\infty}^{\infty} \frac{d c_{q}}{\sqrt{2 \pi \hbar}} \;,
\end{align}
where $\mathcal{N}$ is a constant. 
However, the above treatment of extracting the zero and the negative eigenvalue leads to 
\begin{align}
\prod_{q=0}^{\infty} \int_{-\infty}^{\infty} \frac{d c_{q}}{\sqrt{2 \pi \hbar}} &
\,\, \longrightarrow \,\,\,  
\left[\prod_{q=2}^{\infty} \int_{-\infty}^{\infty} \frac{d \tilde{c}_{q}}{\sqrt{2 \pi \hbar}} \right] \int_{-\beta / 2}^{\beta / 2} \frac{d \tau_{0}}{\sqrt{2\pi \hbar}} \left|\frac{dc_{0}(\tau_{0})}{d\tau_{0}}\right| \int_{0}^{\beta } \frac{d \xi}{\sqrt{2\pi \hbar}} \left|\frac{dc_{1}(\xi)}{d\xi}\right|  \;.
\end{align}
Now we can write for the full partition function element of a single bounce as
\begin{align}
z^{(1)}_{0}
&=
\mathcal{N} \left[ \prod_{q=2}^{\infty} \int_{-\infty}^{\infty} \frac{d \tilde{c}_{q}}{\sqrt{2 \pi \hbar}}  \right]
\int_{-\beta/2}^{\beta/2} \frac{d\tau_0}{\sqrt{2\pi\hbar}} \left|\frac{dc_{0}(\tau_0)}{d\tau_0}\right|
 \int_{0}^{\beta } \frac{d \xi}{\sqrt{2\pi \hbar}} \left|\frac{dc_{1}(\xi)}{d\xi}\right| e^{-\frac{1}{\hbar}S_{cl}[x^{(1)}_{cl}(\tau,\xi)]}\;
 e^{-\frac{1}{\hbar}\frac{1}{2}\sum\limits_{q=2}^{\infty}   \lambda^{(B)}_q \tilde{c}^2_q} \;,
\end{align}
The Jacobian $|dc_{0}(\tau_0)/d\tau_0|$ is provided in Eq.~(\ref{Eq.Jac1}) in which the second part vanishes because it is linear in $\tilde{c}_q$, 
\begin{align}
   \int_{-\infty}^{\infty}d\tilde{c}_q \;\tilde{c}_q\; e^{-\frac{1}{2\hbar}\lambda_q \tilde{c}_q^2}=0
   \quad\quad q \geq 2 
\end{align}
%
%
Therefore we can replace $|dc_{0}(\tau_0)/d\tau_0| \rightarrow 1/A_{\tau_0}$ and use  Eq.~(\ref{Eq.Atau0}).
For the integration over $\xi$ we use the steepest decent method, namely we find the fixed bounce time $\xi_B$ such that
$ {\left. dS(\xi)/d\xi \right|}_{\xi_B} = 0$  leading to
\begin{equation}
z_0^{(1)}=
e^{-\frac{1}{\hbar}S_{cl}[x^{(1)}_{cl}(\tau,\xi_B)]}
\frac{\mathcal{N}   \beta}{A_{\tau_0}\sqrt{2\pi\hbar} }
\frac{1}{\prod_{q=2}^{\infty}\sqrt{\lambda_{q}^{(B)}}}
 \int_{-\infty}^\infty\frac{d\varphi}{\sqrt{2\pi\hbar}} \,\, e^{-\frac{1}{2\hbar}\lambda_1^{(B)} \varphi^2 
}\; ,
\label{Eq.singlebounce}
\end{equation}
in which the last integral corresponds to the contribution of the breathing mode around $\xi_B$ with $\lambda_1^{(B)}<0$.
We denote $|\lambda_1^{(B)}|=-\lambda_1^{(B)}>0$. 
As a consequence the integral of Eq.~(\ref{Eq.singlebounce}) diverges. 
This is not surprising as we want to calculate the eigenvalues for a system that is metastable at the position $x=0$ [\onlinecite{Callan1977}]. 
The integral has to be analytically continued to avoid this divergence, 
as discussed in chapter 17 of the book of Kleinert [\onlinecite{Kleinert1995}],  in the paper of Langer [\onlinecite{Langer1967}] and the one of Callan and Coleman
[\onlinecite{Callan1977}]. 
The basic idea is to deform a stable action into a metastable one by keeping track of the eigenvalue $\lambda_1^{(B)}$ leading to
\begin{equation}
\frac{1}{\sqrt{2\pi\hbar}}\int d\varphi \,\,  e^{\frac{1}{2\hbar}|\lambda_1^{(B)}| \varphi^2}=\frac{1}{2}\frac{i}{\sqrt{|\lambda_1^{(B)}|}}
\;,
\end{equation}
and we obtain the imaginary part previously discussed. 
Finally, the single bounce amplitude reads
\begin{align}
z_L^{(1)} 
& =e^{-\frac{1}{\hbar}S_{cl}[x^{(1)}_{cl}(\tau,\xi_B)]}
\frac{\mathcal{N}   \beta}{A_{\tau_0}\sqrt{2\pi\hbar} }
\frac{1}{\prod_{q=2}^{\infty}\sqrt{\lambda_{q}^{(B)}}}
\left( \frac{i}{ 2\sqrt{|\lambda_1^{(B)}|}} \right) 
\nonumber \\
& =\frac{i \beta}{2} e^{-\frac{1}{\hbar}S_{cl}[x^{(1)}_{cl}(\tau,\xi_B)]}
\frac{1}{A_{\tau_0}\sqrt{2\pi\hbar} }  \left( \frac{ \mathcal{N}  }{  \prod_{q=0}^{\infty}\sqrt{\lambda_{q}^{(0)}} } \right)
\frac{ \prod_{q=0}^{\infty}\sqrt{\lambda_{q}^{(0)}}  }{\prod_{q=2}^{\infty}\sqrt{\lambda_{q}^{(B)}}}
\left( \frac{1}{ \sqrt{|\lambda_1^{(B)}|}} \right) 
\end{align}
where we have introduced the product of the eigenvalues for the harmonic potential associated to the metastable well with frequency $\omega_0$.
Then one can notice that
\begin{align}
\frac{ \mathcal{N} } 
{    \prod_{q=0}^{\infty} \sqrt{ \lambda_{q}^{(0)} }  }  
&
= 
Z_0^{(0)}
= \frac{1}{\sqrt{2\pi {\langle x^2 \rangle} }} e^{- \frac{\beta E_{GS}}{2}  }
\end{align}
corresponds to the value of the amplitude for the case of a harmonic potential in presence of dissipation,
with $ E_{GS}$ the ground state energy and  $ {\langle x^2 \rangle} $ the harmonic fluctuations.
We also set 
\begin{align}
K  =  
\frac{1}{ \sqrt{2\pi \hbar} }  \frac{1}{  A_{\tau_0}}
\, 
\sqrt{ \frac{ \prod_{q=0}^{\infty} \lambda_{q}^{(0)}    } {  \prod_{q=2}^{\infty}  \lambda_{q}^{(B)}   } }
\frac{1}{\sqrt{|\lambda_1^{(B)}|}} 
\, .
\label{Eq.prefactormeta}
\end{align}
containing the ratio of determinants, the Jacobian prefactor $A_{\tau_0}$ and the negative eigenvalue.
The final formula reads
\begin{align}
z_0^{(1)} 
& 
= \frac{i\beta}{2}
Z_{0}^{(0)} K e^{-\frac{1}{\hbar}S_{cl}[x^{(1)}_{cl}(\tau,\xi_B)]}
\;.\label{Eq.prefactor}
\end{align}
Recalling Eq.~(\ref{eq:sum}), we must sum over many bounce paths.
In the zero temperature limit we use the dilute gas of bounces in which the different  bounces do not interact with each other
and $S_{cl}[x_{cl}^{(n)} (\tau) ] \simeq \, n \,\, S_{cl}[x_{cl}^{(1)} (\tau) ]$ and, in a similar way, the integral over the fluctuations paths $\delta x(\tau)$

\begin{align}
Z_{0} 
& =
Z_{0}^{(0)}
\sum_{n=0}^{\infty}
\frac{1}{ n! } 
{\left( \frac{i \beta}{2} \right)}^n
K^n
e^{-n \,\, \frac{1}{\hbar}S_{cl}[x_{cl}^{(1)}(\tau,\xi_B)] }  \, .
\label{eq:K-sum}
\end{align}
Assuming that the amplitude decays exponentially as
\begin{equation}
Z_{0} 
=
\frac{1}{\sqrt{2\pi\langle x^2\rangle}}e^{-\frac{\beta}{2 }\left( \frac{E_{GS}}{\hbar} -i  \Gamma \right)}
\,,
\label{eq:heuristic}
\end{equation}
we finally find by comparison between Eq.~(\ref{eq:K-sum}) and Eq.~(\ref{eq:heuristic})
\begin{equation}
\Gamma = K  \, e^{-\frac{1}{\hbar}S_{cl}} 
 \, .
\label{Eq.Gammadecay}
\end{equation}

%
%
%
%
%
\section{The bounce path and the classical action in presence of position and momentum dissipation}\label{App:B}

\subsection{General integral formulas}
The parametrization of the metastable potential displayed in Fig.~\ref{Fig1-supp}a reads 
\begin{align}
V(x)=
\left\{
\begin{array}{lr}
\frac{1}{2} m \omega_0^2 x^2 & \,\,\,\,  x < a \\
\frac{1}{2} m \omega_0^2 {\left( x - x_m \right)}^2 - \Sigma& \,\, \,\,  a< x < x_m\\
-\Sigma &  \,\, \,\,  x_m < x\\
\end{array}
\right.
\, .
\end{align}
We set $V_0=m\omega_0^2 a^2/2$  and apply the 
 matching condition  $m \omega_0^2 {\left( a - x_m \right)}^2 /2 - \Sigma = V_0$. We obtain 
$x_m= a \left[ 1+ \sqrt{1+\Sigma/V_0} \right]$ 
for the point where the potential becomes flat.

We use the Matsubara frequency decomposition for the classical bounce path that minimizes 
the action $S =S_0+S_{dis}$ defined in the main text. 
Hereafter we use the notation  $x^{(1)}_{cl}(\tau,\xi)\equiv x_{cl}(\tau,\xi)$
\begin{equation}
x_{cl}(\tau,\xi)=\frac{1}{\beta}\sum_{l=-\infty}^{\infty}x_{l}(\xi)e^{i\omega_{l}\tau}
\label{Eq.bouncepathmeta}\; ,
\end{equation}
with the Matsubara frequency $\omega_{l}=2\pi l/\beta$ ($l$ integer) 
and $\xi$ is the bounce time as defined in the previous section.
From the condition $dS/dx_{l}=0$, we obtain the solution 
\begin{align}
x_{l}(\xi)
=
\frac{2\omega_{0}^{2}a\left(1+\sqrt{1+\Sigma/V_0}\right)\sin(\omega_{l}\frac{\xi}{2})}{\omega_{l}\left(\omega_{l}^{2}+\omega_{0}^{2}+\frac{F^{(x)}_{l}}{m}+\omega_{l}^{2}\frac{F^{(p)}_{l}}{m}\right)}
\;. \label{Eq.xlmeta}
\end{align} 
Inserted in Eq.~(\ref{Eq.bouncepathmeta}) the path solves the equation of motion
\begin{align}
-m\frac{d^{2}}{d\tau^{2}}x_{cl}(\tau,\xi)+\frac{d{V(x(\tau))}}{dx}\left|_{x_{cl}}\right.+\int_{-\frac{\beta}{2}}^{\frac{\beta}{2}}d\tau'F^{(x)}(\tau-\tau')x_{cl}(\tau',\xi)-\int_{-\frac{\beta}{2}}^{\frac{\beta}{2}}d\tau'F^{(p)}(\tau-\tau')\frac{d^{2}}{d\tau'^{2}}x_{cl}(\tau',\xi)%
=0\;,\label{Eq.equationofmotion}
\end{align}
where we see that the kernel for momentum dissipation couples to the acceleration, as mentioned in the main text. 

Then, we insert the path into the action $S$ and use the condition $dS_{cl}/d\xi = 0$  to
find the  relation determining the saddle point of the bounce time, which we denoted $\xi_B$, in presence of both dissipative couplings. 
In the limit $\beta\rightarrow\infty$, the quantity  $\xi_B$ solves  the following integral equation
\begin{align}
\frac{1}{\pi}
\int_{0}^{\infty} \frac{d\omega}{\omega}
\frac{ \omega_0^2 \sin(\omega\xi_{B}) }{ \left( \frac{\omega^{2}}{1+\tau_{p}\omega f_c(\omega)} + \omega_{0}^{2} + \gamma\omega f_{c}(\omega) \right) }
=
\frac{1}{\sqrt{1+\frac{\Sigma}{V_0}}+1}
\label{Eq.timeintegraldis}
\,.
\end{align}
The integral equation determining the action reads
\begin{align}
S_{cl} &
 =-\frac{2\omega_{0}^{2}V_0\left(\sqrt{1+\frac{\Sigma}{V_0}}+1\right)^{2}}{\pi}
 \int_{0}^{\infty}d\omega\frac{\left(1-\cos\left(\omega\xi_B\right)\right)}{\omega^{2}
 \left(\frac{\omega^{2}}{1+\tau_{p}\omega f_c(\omega)}+\omega_{0}^{2}+\gamma\omega f_{c}(\omega)\right)}
 +
 2V_0 \left( 1+\sqrt{1+\frac{\Sigma}{V_0}}\right) \xi_{B}\;.
\label{Eq.Actiongeneral}
\end{align}
The Eqs.~(\ref{Eq.timeintegraldis}) and (\ref{Eq.Actiongeneral}) for $\xi_{B}$ and $S_{cl}$ correspond to the  equations reported in the main text, which we computed numerically 
in the general case.

Examples of the bounce path are shown in Fig.~\ref{Fig.Fig12-supp} for  two fixed values of $\Sigma$.
The dissipative interaction determines the path trajectory, in particular the bounce time $\xi_B$ and the returning point
$x_{esc}$.
In Fig.~\ref{Fig.Fig12-supp},  
the black solid line corresponds to the non dissipative case discussed in the main text, 
the blue dashed line shows the influence of position dissipation, 
the red dotted line of the momentum dissipation and the green dotted-dashed line to both dissipative interactions.
While the bounce becomes wider in presence of pure position dissipation, it shrinks for pure momentum dissipation.
However, in both cases, the returning point $x_{esc}$ shifts to larger values. 
When both dissipative couplings are present, for $\Sigma=V_0$ in Fig.~\ref{Fig.Fig12-supp}, 
the bounce is similar to the one of position dissipation only.
By increasing $\Sigma$ this behavior changes: in Fig.~\ref{Fig.Fig12-supp}b the bounce almost coincides with the one of momentum dissipation. 
Hence, there is a crossover in the influence of the two different dissipative interactions which depends on the steepness of the potential. 
\begin{figure}[t]
	\centering
	\includegraphics[width=0.8\linewidth]{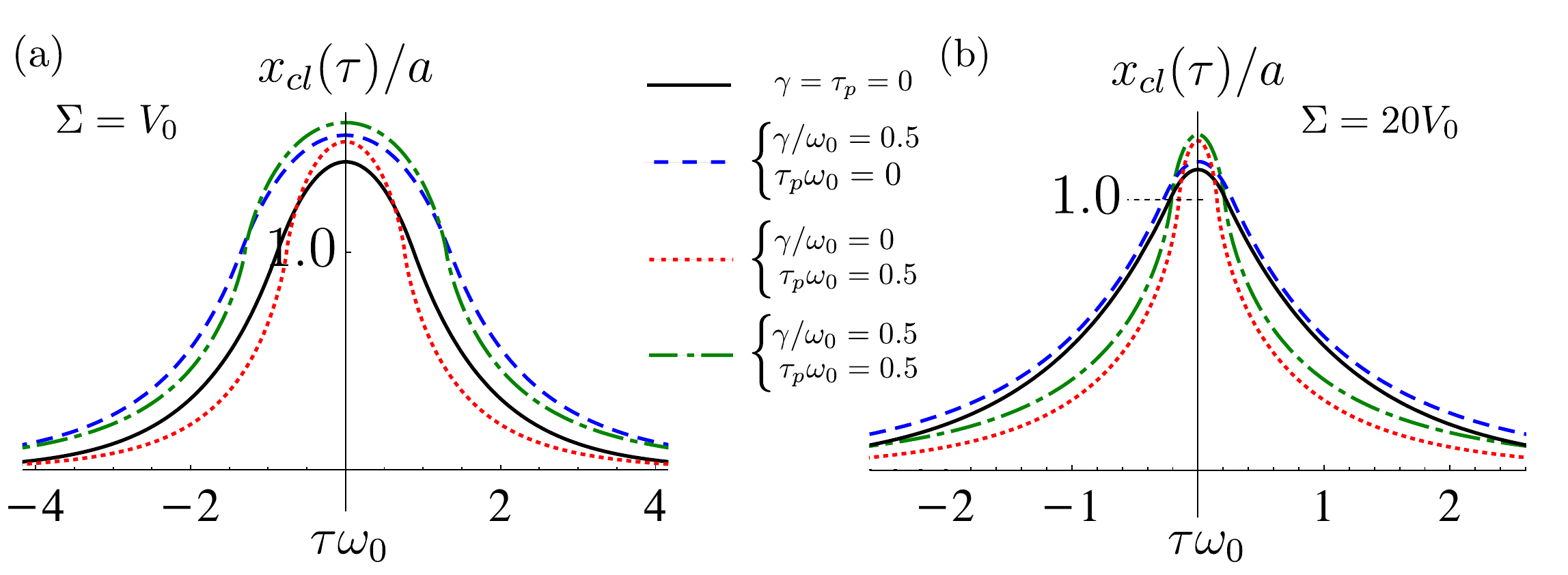}
	\caption{Example of bounce paths with the two dissipative couplings for (a) $\Sigma=V_0$ and (b)  $\Sigma=20V_0$. 
	The returning point $x_{esc}$ shifts to higher values in both dissipative cases.}
	\label{Fig.Fig12-supp}
\end{figure}

\subsection{Analytical formulas and expansions}
All analytical formulas are calculated in the limit $\omega_c\gg\gamma,\tau_p\omega_0^2,\omega_0$ meaning that the cutoff $\omega_c$ is the largest frequency in the problem.
We will see that, while for pure conventional dissipation the cutoff is irrelevant in the limit $\omega_c \gg \gamma,\omega_0$,  it plays an important role in the case of momentum dissipation. 

First, we consider Eq.~(\ref{Eq.timeintegraldis}) and expand the integral using the roots of the denominator. 
This yields the exact result
\begin{align}
\frac{1}{\pi}\int_{0}^{\infty}d\omega\frac{\sin(\omega\xi_{B})}{\omega\left(\frac{\omega^{2}}{1+\tau_{p}\omega f_c(\omega)}+\omega_{0}^{2}+\gamma\omega f_{c}(\omega)\right)}=\frac{1}{2}\frac{1}{\omega_{0}^{2}}+\frac{1}{\pi z\omega_{c}^{2}}
\sum_{i=1}^{4} \,\, \mathcal{T}_{i} \,\, f[\xi_{B}\omega_{c}\Lambda_{i}]
\; , \label{Eq.timeana}
\end{align}
where $ 1/z=\left(1+\tau_{p}\omega_{c}\right)$ and the coefficients
\begin{align}
\mathcal{T}_{i}&=\frac{z-\Lambda_{i}\left(z+1\right)+\Lambda_{i}^{2}}{\Lambda_{i}(\Lambda_{i}-\Lambda_{j})(\Lambda_{i}-\Lambda_{k})(\Lambda_{i}-\Lambda_{l})}\quad\text{with $(j,k,l)\neq i$} 
\end{align}
for $i,j,k,l=1,2,3,4$, whereas the auxiliary function is defined as
\begin{align}
f(zx)&=Ci(zx)\sin(zx)+\frac{1}{2}\cos(zx)(\pi-2Si(zx))=\int_{0}^{\infty}du\frac{\sin(z\;u)}{(x+u)}\;.
\end{align}
The roots of the denominator in the limit $\omega_c\gg\gamma,\tau_p\omega_0^2,\omega_0$ are the same as in [\onlinecite{Rastelli2016ge}]
\begin{align}
\Lambda_{1,2}&=\frac{1}{(1+\sigma^{2})}\frac{\omega_0}{\omega_c}\left(P_{+}\pm\sqrt{ P_-^{2}-1}\right)\nonumber \\\Lambda_{3,4}&=1-\frac{\omega_0}{\omega_c}\frac{P_{+}}{(1+\sigma^{2})}\pm i\sigma\left(1+\frac{\omega_0}{\omega_c}\frac{P_{+}}{(1+\sigma^{2})}\right)\;, \label{Eq.rootsmeta}
\end{align}
with
\begin{align}
\sigma^{2}=\gamma\tau_{p},\quad P_{+}=\frac{\gamma+\tau_{p}\omega_{0}^{2}}{2\omega_{0}}\quad \text{and}\quad P_{-}=\frac{\gamma-\tau_{p}\omega_{0}^{2}}{2\omega_{0}}\;.  
\end{align}
In the limit $\xi_{B}\omega_{0}\ll1$ and $P^2_->1$, we find 
\begin{align}
\xi_{B}\omega_{c}\Lambda_{1,2}&=\xi_{B}\omega_{0}\frac{1}{(1+\sigma^{2})}\left(P_{+}\pm\sqrt{ P_-^{2}-1}\right)
\ll1 
\label{Eq.regions}
\end{align}
For the case $P^2_-<1$  we analogously obtain
\begin{align}
\xi_{B}\omega_{c}\Lambda_{1,2}&=\xi_{B}\omega_{0}\frac{1}{(1+\sigma^{2})}\left(P_{+}\pm i\sqrt{| P_-^{2}-1|}\right)\; ,
\label{Eq.regions1-1}
\end{align}
and find in the limit $\xi_{B}\omega_{0}\ll1$ also $\text{Im}\left(\xi_{B}\omega_{c}\Lambda_{1,2}\right)\ll1$ and $\text{Re}\left(\xi_{B}\omega_{c}\Lambda_{1,2}\right){\ll1}$.
For the other two roots, we obtain
\begin{align}
\xi_{B}\omega_{c}\Lambda_{3,4}
=&
\;\xi_{B}\omega_{c}\left(1\pm i\sigma-\frac{\omega_0}{\omega_c}\frac{P_{+}}{(1+\sigma^{2})}\pm \frac{\omega_0}{\omega_c}i\sigma\frac{P_{+}}{(1+\sigma^{2})}\right)\nonumber\\&\!\!\!\!\!\!\!\!\!\!\!\!\stackrel{\xi_{B}\omega_{0}\ll1}{\approx}\xi_{B}\omega_{c}(1\pm i\sigma)=
\begin{cases}
\text{Re}(\xi_{B}\omega_{c}(1\pm i\sigma))\ll1  ,\,\,  \text{Im}(\xi_{B}\omega_{c}(1\pm i\sigma))\ll1 & \mbox{for}\,\,\,  \xi_{B}\omega_{c}\ll1\\
\text{Re}(\xi_{B}\omega_{c}(1\pm i\sigma))\gg1 , \,\, \text{Im}(\xi_{B}\omega_{c}(1\pm i\sigma))\gg1  & \mbox{for} \,\,\,  \xi_{B}\omega_{c}\gg1 
\label{Eq.regions2}
\end{cases}
\end{align}
Because $\xi_B$ strongly depends on the shape of the potential, the different regimes in Eq.~(\ref{Eq.regions2}) depend on $\Sigma$ and the cutoff $\omega_c$.

We conclude this section giving an analytic  formula for Eq.~(\ref{Eq.Actiongeneral}). 
The integral can be expanded similarly via the same roots as in Eq.~(\ref{Eq.rootsmeta}), yielding the action
\begin{align}
S_{cl}=\frac{2V_0\left(\sqrt{1+\frac{\Sigma}{V_0}}+1\right)^{2}}{\omega_0\pi}\left(\frac{\gamma}{\omega_{0}}\left(C+\log(\xi_{B}\omega_{0})\right)-\frac{\omega_0^3}{z\omega_{c}^{3}}\sum_{i=1}^{4}\frac{\mathcal{T}_{i}}{\Lambda_{i}}\left(\ln\left(\frac{\Lambda_{i}\omega_c}{\omega_{0}}\right)+g[\omega_{c}\xi_{B}\Lambda_{i}]\right)\right)-\xi_{B}\Sigma\;, \label{Eq.anaaction}
\end{align}
where the auxiliary function $g$ is defined as
 \begin{align}
g(zx)=&-\cos(zx)Ci(zx)-\sin(zx)Si(zx)+\frac{1}{2}\sin(zx)\pi =\int_{0}^{\infty}du\frac{\cos\left(z\; u\right)}{(x+u)}\;.
\end{align}
We see that the arguments of $g(x)$ in Eq.~(\ref{Eq.anaaction}) are the same as the ones of $f(x)$ in Eq.~(\ref{Eq.timeana}) 
and therefore the limits defined in Eqs.~(\ref{Eq.regions}) -- (\ref{Eq.regions2}) can be used.
In the following, we will use the approximated solution for the roots  to expand the auxiliary functions $g(x)$ and $f(x)$ for the analytical formulas for the action.

\subsubsection{The limit $\xi_B\omega_0\ll 1$ and  $\xi_B\omega_c\ll 1$} 
In the case of $\Sigma \gg V_0$, the bounce time becomes almost zero and the condition $\xi_B\omega_0\ll 1$ is always verified as $\xi_B \rightarrow 0$.
Moreover, at fixed $\omega_c$,  increasing $\Sigma$ eventually leads to $\xi_B\omega_c\ll 1$.
Then we can expand $g(x)$ and $f(x)$ for small arguments and we find
\begin{align}
\xi_{B}\omega_{0}&=\frac{\hbar}{m\omega_0}\frac{1}{\sqrt{1+\frac{\Sigma}{V_0}}+1}\frac{1}{\langle x^{2}\rangle}\;, \label{Eq.timefinal1}
\end{align}
where we introduced the quantum fluctuations in presence of both dissipative couplings [\onlinecite{Rastelli2016ge}]
\begin{equation}
\langle x^{2}\rangle=\langle x^{2}\rangle_{0}\frac{2}{\pi\left(1+\sigma^{2}\right)}\left[\tau_p \omega_0\left(\ln\left(\frac{\omega_{\mathrm{c}}}{\omega_{0}}\right)+\sigma\arctan(\sigma)+\ln\left(1+\sigma^{2}\right)\right)+\frac{\left(1+\tau_p \omega_0 P_-\right)}{\sqrt{\left|1- P_-^{2}\right|}}\Theta_{q,p}\right] \label{Eq.fluctuationsfinal}\; ,
\end{equation}
where
\begin{equation}
\Theta_{q,p}=\left\{ \begin{array}{ll}
\arctan(\sqrt{1- P_-^{2}}/P_+)\text{ for } & \left|P_-\right|<1\\
\operatorname{arctanh}(\sqrt{ P_-^{2}-1}/P_+)\text{ for } & \left| P_-\right|>1
\end{array}\right. \;.
\end{equation}
Here $\langle x^2 \rangle_0 $ are the harmonic quantum fluctuations in the well without dissipation.
This result we also find by simply expanding the numerator $\sin(\xi \omega)\approx \xi\omega$ of the integrand in Eq.~(\ref{Eq.timeintegraldis}), 
since it is peaked around $\omega \approx 0$ as can be seen from the denominator. 
From a such expansion, we find the same integral as in [\onlinecite{Rastelli2016ge}] 
from which we  identify the quantum fluctuations at zero temperature in presence of dissipation. 
We can analogously expand the action in Eq.~(\ref{Eq.anaaction}) and find by inserting Eq.~(\ref{Eq.timefinal1})
\begin{align}
S_{cl}=\frac{\hbar}{2}\frac{ a^{2}}{\langle x^{2}\rangle} \label{Eq.actionana1} \;.
\end{align}
This result coincides with the action of the potential with an infinitely sharp barrier, as expected in the limit $\Sigma\rightarrow \infty$. 
Expanding Eq.~(\ref{Eq.actionana1}) in the pure momentum dissipative case to first order in $\tau_p\omega_0$ we find Eq.~(7) of the main text. 
The cutoff frequency $\omega_c$ in $\langle x^2 \rangle$ is related to the environment coupled to the momentum, 
while the effect of the cutoff frequency for the position bath drops out of the calculation in the limit $\omega_c\gg \gamma,\omega_0$.

\subsubsection{The limit $\xi_B\omega_0\ll 1 \ll \xi_B\omega_c$ } 
Since we are in the regime $\omega_c \gg \omega_0$, there exists a regime $\xi_B \omega_0  \ll 1 \ll \xi_B \omega_c$, 
which we may reach by reducing $\Sigma$ with respect to the case discussed above (note that $\Sigma \gg V_0$ still holds). 
In this limit, we can expand the functions $f[\xi_B\omega_c\Lambda_{1,2} ]$ and $g[\xi_B\omega_c\Lambda_{1,2} ]$ 
for small arguments, but $f[\xi_B\omega_c\Lambda_{3,4} ]$ and $g[\xi_B\omega_c\Lambda_{3,4} ]$ for large arguments 
(see  Eqs.~(\ref{Eq.regions}), (\ref{Eq.regions2})). 

We find for the bounce time equation 
\begin{align}
\frac{1}{\pi}\frac{\xi_{B}\omega_{0}}{(1+\sigma^{2})}\left[\frac{\left(1+\tau_{p}\omega_{0} P_{-}\right)}{2\sqrt{ P_-^{2}-1}}\ln\left(\frac{\Lambda_1}{\Lambda_{2}}\right)-\tau_{p}\omega_{0}\left(\ln(\omega_{0}\xi_{B})-\frac{1}{2}\ln(1+\sigma^{2})+(C-1)\right)\right]
=
\frac{1}{\sqrt{1+\frac{\Sigma}{V_0}}+1}\;, \label{Eq.timefinal2}
\end{align}
which is, in presence of momentum dissipation, a nonlinear equation to be solved numerically. 
In the limit here discussed, the action reads
\begin{align}
S_{cl}=\frac{V_0}{\omega_0}\left(\sqrt{1+\frac{\Sigma}{V_0}}+1\right)
\xi_{B}\omega_{0}-\frac{V_0}{\omega_0}\frac{\left(\sqrt{\frac{\Sigma}{V_0}+1}+1\right)^{2}}{2\pi(1+\sigma^{2})}\xi_{B}^{2}\omega_{0}^{2}\tau_{p}\omega_{0}\;. 
\label{Eq.actionana2} 
\end{align}
Hence, by inserting Eq.~(\ref{Eq.timefinal2}) into Eq.~(\ref{Eq.actionana2}),  
the action depends logarithmically on the  bounce time $\xi_B$ and therefore on $\Sigma$. 
We also see, that this dependence vanishes in the limit $\tau_p=0$, meaning that momentum dissipation induces a stronger dependence 
on the shape of the barrier as discussed in the main text. 
The combined Eqs.~(\ref{Eq.timefinal2}) and (\ref{Eq.actionana2}) are displayed via the dashed lines in Fig.~2b of the main text. Further, for the case $\gamma=0$ expanding (\ref{Eq.timefinal2}) and (\ref{Eq.actionana2}) to first order in $\tau_p\omega_0$ leads to Eq.~(8) in the main text.

\subsubsection{The limit $\xi_B\omega_0\gg 1$ and  $\xi_B\omega_c\gg 1$} 
Finally we analyze the regime $\xi_B\omega_0\gg 1$ and  $\xi_B\omega_c\gg 1$ which is reached for $\Sigma\ll V_0$. 
The action can then be approximated as
\begin{equation}
    S_{cl}  \approx \frac{\epsilon_{0}}{ \omega_{0}}+\frac{8}{\pi} \frac{V_{0}}{ \omega_{0}} \frac{\gamma}{\omega_{0}} \ln \left(\xi_{B} \omega_{0}\right)- \Sigma \xi_{B}\;,
\end{equation}
where $\xi_{B} \omega_{0} \approx \frac{8}{\pi} \frac{V_0}{\Sigma} \frac{\gamma}{\omega_{0}}$ and
\begin{equation}
    \frac{\epsilon_{0}}{ \omega_{0}}=\frac{8}{\pi} \frac{V_{0}}{ \omega_{0}}\left[\frac{\gamma}{\omega_{0}}\left(C-\frac{\ln \left(1+\sigma^{2}\right)}{2}\right)-\frac{\left(\frac{\gamma}{\omega_{0}} P_{-}-1\right)}{2 \sqrt{P_{-}^{2}-1}} \ln \left(\frac{\Lambda_1}{\Lambda_{2}}\right)\right]\;,
\end{equation}
with $C$ the Euler constant.
In particular, in the limit $\gamma=0$ and $\tau_p\omega_0 \ll 1$ we find
\begin{equation}
    S_{c l}  \approx \frac{4 V_{0}}{ \omega_{0}}-\frac{4}{ \pi} V_{0} \tau_{p}
\end{equation}
yielding Eq.~(9) for $\mathcal{E}$ in the main text. 
Note that for $\tau_p=0$ and $\gamma \neq 0$  we recover the action for the incoherent decay in a (slightly) asymmetric parabolic double well in presence of dissipation 
 [\onlinecite{Grabert1987}] \footnote{Note that for $\Sigma \rightarrow 0$ the steepest decent approximation is no longer valid, because the negative eigenvalue $\lambda_1^{(B)}$ approaches zero. However, for the minimal asymmetry parameter $\Sigma =0.01 V_0$, we use in this paper, the negative eigenvalue is $|\lambda_1^{(B)}/(m\omega_0^2)|\approx 0.01$ and the approximation still justified}. 
%
In particular, within the steepest decent approximation, the quantity $\mathcal{E}$ can be approximated as
\begin{equation}
 \mathcal{E}_{\tau_{p}=0}^{\Sigma\ll V_{0}}\approx e^{-\frac{\epsilon_{1}(\gamma)}{\hbar\omega_{0}}}\left(\frac{8}{\pi}\frac{\gamma}{\omega_{0}}\frac{V_{0}}{\Sigma}\right)^{-\frac{8}{\pi}\frac{V_{0}}{\hbar\omega_{0}}\frac{\gamma}{\omega_{0}}}\;, \label{Eq.posshape}
\end{equation}
with
\begin{equation}
    \frac{\epsilon_{1}(\tau_p=0)}{\hbar\omega_{0}}=\frac{8}{\pi}\frac{V_{0}}{\hbar\omega_{0}}\frac{\gamma}{\omega_{0}}(C-1)-\frac{4V_{0}}{\hbar\omega_{0}}\left(\frac{2}{\pi}\frac{\left(\frac{\gamma}{\omega_{0}}P_{-}-1\right)}{2\sqrt{P_{-}^{2}-1}}\ln\left(\frac{\Lambda_1}{\Lambda_{2}}\right)+1\right)\;,
\end{equation}
and the decay is exponentially suppressed in presence of pure position dissipation. 
The Eq.~(\ref{Eq.posshape})  shows that  $\mathcal{E}$ depends on $\Sigma$ only via the prefactor of the exponential function and is independent of the cutoff $\omega_c$.

\section{The prefactor K}\label{App:C}
In this section we present an overview of the calculation of the prefactor $K$, 
defined in Eq.~(\ref{Eq.prefactormeta}), for the potential shown in Fig.~\ref{Fig1-supp}a. 
For a more detailed introduction we refer to Ref.~[\onlinecite{Grabert1987}].
Example of results for the prefactor $K$ are reported in Fig.~\ref{FigK}d in which $K$ is scaled with its value in absence of dissipation $K_0$.
Similarly to the exponential function,  $K$ is enhanced in presence of momentum dissipation.

\subsection{The ratio between the determinants}
We start by calculating the ratio of determinants defined in Eq.~(\ref{Eq.prefactormeta}), namely the ratio between by the two products of the two sets of eigenvalues
\begin{align}
R
=\left(\frac{\prod_{q=0}^{\infty}\sqrt{\lambda_{q}^{(0)}}}{\prod_{q=2}^{\infty}\sqrt{\lambda_{q}^{(B)}}}\right)\;.
\end{align}

The eigenvalues of the bounce path $\lambda_q^{(B)}$ are defined via Eq.~(\ref{eq:schr}), while the eigenvalues $\lambda_q^{(0)}$ are associated to the following equation.
\begin{align}
\left(-m\frac{d^{2}}{d\tau^{2}}
+m\omega_0^2\right)
y^{(0)}_{q}(\tau)
+\int_{-\frac{\beta}{2}}^{\frac{\beta}{2}}d\tau'F^{(x)}(\tau-\tau')y^{(0)}_{q}(\tau')
-\int_{-\frac{\beta}{2}}^{\frac{\beta}{2}}d\tau'F^{(p)}(\tau-\tau')\frac{d^{2}}{d\tau'^{2}}y^{(0)}_{q}(\tau')
=\lambda_{q}^{(0)}y^{(0)}_{q}(\tau)
\, . \label{eq:schrmeta}
\end{align}
Taking the second derivative of the potential in Eq.~(\ref{eq:schr}) along a bounce trajectory yields $\mathcal{V}[x_{cl}^{(1)}]=\frac{d^2V[x(\tau)]}{dx^2}\left|_{x_{cl}^{(1)}}\right.$.
Assuming that the time $\tau$ is the space and $y(\tau)$ the wavefunction, in absence of dissipation, the equation corresponds to the  
Schr{\"o}dinger  equation with a potential at constant value $m\omega^2_0 $ containing two delta-potential wells (at the times when the periodic path crosses the discontinuity at $x=a$).
Each well has one bound state and the finite size of the bounce (determined by $\xi_B$) leads to a hybridization of the two wells yielding the two bound states $\lambda_0^{(B)}$ and $\lambda_1^{(B)}$ (denoting the zero mode due to translational invariance and the negative eigenvalue of the breathing mode). 
The rest of the eigenvalues forms a continuum above $m\omega_0^2$. 
\mbox{}\\

We rewrite the ratio of the determinants containing only the continuum eigenvalues defined in Eq.~(\ref{Eq.prefactormeta}) via 
\begin{align}
R=\left(\frac{\prod_{q=0}^{\infty}\sqrt{\lambda_{q}^{(0)}}}{\prod_{q=2}^{\infty}\sqrt{\lambda_{q}^{(B)}}}\right)=e^{\frac{1}{2}\int_{m\omega_{0}^{2}}^{\infty}d\lambda\ln\left(\lambda\right)(\rho_0(\lambda)-\rho(\lambda))}
\;, \label{Eq.factorR}
\end{align} 
where we defined the spectral densities 
$\rho(\lambda)=\sum_{q=2}^{\infty} \delta\left(\lambda_{q}^{(B)}-\lambda\right)$ and $\rho_0(\lambda)=\sum_{q=0}^{\infty} \delta\left(\lambda_{q}^{(0)}-\lambda\right)$.
We rewrite the spectral densities using the (retarded) Greens functions of the respective problem 
\begin{align}
\rho_{0}(\lambda) & =\frac{1}{\pi}\text{Im}\left(G_{\lambda}^{(0)}(0)\right), \quad \text{and} \quad \rho(\lambda)=\frac{1}{\pi}\text{Im}\left(G^{(B)}_{\lambda}(0,0)\right)\label{Eq.Green1}\;,
\end{align}
in which $G_\lambda^{(0)}$ is given by
\begin{align}
G_{\lambda}^{(0)}(\tau)
&=\frac{1}{\pi m}\int_{0}^{\infty}d\omega\frac{\cos(\omega\tau)}{\frac{\omega^{2}}{1+\tau_{p}\omega f_c(\omega)}+\omega_{0}^{2}+\gamma\omega -\frac{\lambda}{m}-i\epsilon}\;,\label{Eq.metagreendis}
\end{align}
with $\epsilon \rightarrow 0$. 
For simplicity we do not consider a high frequency cutoff for the environment coupled to the position as it is irrelevant. By contrast, 
the cutoff for the momentum bath has to be finite, otherwise $R$ diverges in the limit $\omega^{(p)}_c\rightarrow \infty$. 
Further, we determine $G^{(B)}_{\lambda}$ in terms of $G^{(0)}_\lambda$ via the Lippmann-Schwinger equation 
\begin{equation}
G^{(B)}_{\lambda}(\tau,\tau'')=G_\lambda^{(0)}(\tau-\tau'')-\int_{-\infty}^{\infty}d\tau'G_{\lambda}^{(0)}(\tau-\tau')\mathcal{V}(x^{(1)}_{cl}(\tau'))G^{(B)}_{\lambda}(\tau',\tau'')
\;. \label{Eq.metagreen}
\end{equation}
Following the calculation outlined in [\onlinecite{Maile2020,Grabert1987}] we find 
\begin{align}
\ln(R)=\frac{1}{2\pi}\left(\left[\ln\left(\lambda\right)\left(\phi^{+}_\lambda+\phi^{-}_\lambda\right)\right]_{m\omega_{0}^{2}}^{\infty}-\int_{m\omega_{0}^{2}}^{\infty}d\lambda\frac{1}{\lambda}\left(\phi^{+}_\lambda+\phi^{-}_\lambda\right)\right)
 \label{Eq.lnR}\;,
\end{align}
with $\phi_\lambda ^\pm=\arg\left(U^{-1}-\left(G_{\lambda}^{(0)}(0)\pm G_{\lambda}^{(0)}(\xi_B)\right)\right) $ 
and $U=m\omega_{0}^{2}\left( a+x_{m}\right)/|\dot{x}_{cl}(\frac{\xi_{B}}{2})|$. 
The phases satisfy $\phi_{m\omega_0^2}^{\pm}=-\pi$  and $\lim\limits_{\lambda\rightarrow \infty}\phi_\lambda^{\pm}=0$ (calculated in App.~\ref{subsec.phases}) leading to the result
\begin{align}
\ln(R_B)&=\ln\left(m\omega_0^2\right) -\frac{1}{2\pi}\int_{m\omega_{0}^{2}}^{\infty}d\lambda\frac{1}{\lambda}\left(\phi^{+}_\lambda+\phi^{-}_\lambda\right)\;. \label{Eq.lnRfinal}
\end{align}
We display the numerical results for $R$ rescaled with the value $R_0$ in absence of dissipation
for different values of $\Sigma$ in Fig.~\ref{FigK}a and as function of the dissipative coupling strength.

\subsection{Determination of the negative eigenvalue $\lambda_B$}
%
We use the above Greens function to obtain the solution for the negative eigenvalue 
via the poles of the T-matrix defined through the Lippmann Schwinger Eq.~(\ref{Eq.metagreen}), see Ref.[\onlinecite{Grabert1987}]. 
The negative eigenvalue is determined by the equation
\begin{align}
|\dot{x}_{cl}(\frac{\xi_{B}}{2})|-m\omega_{0}^{2}a\left(1+\sqrt{1+\frac{\Sigma}{V_0}}\right)\left(G_{-|\lambda_1^{(B)}|}^{(0)}(\xi_{B})+G_{-|\lambda_1^{(B)}|}^{(0)}(0)\right)
=0
\; .
\label{Eq.negativeconv}
\end{align}
In the non dissipative case, we can solve the integral (\ref{Eq.metagreendis}) leading to the following equation for the negative eigenvalue 
which only depends on the bounce path via the bounce time $\xi_B^{(0)}$
\begin{align}
\frac{1}{2\omega_{0}}\left(1-e^{-\omega_{0} \xi^{(0)}_{B} }\right)-\frac{1+e^{- \xi_B^{(0)} \sqrt{ \omega_{0}^{2}+\frac{|\lambda_1^{(B,0)}|}{m} }  } }{2\sqrt{\omega_{0}^{2}+\frac{|\lambda_1^{(B,0)}|}{m}}}
=0
\; .
\end{align}
In Fig.~\ref{FigK}c we show numerical results for the value $\lambda_1^{(B)}$ in presence of both dissipative couplings, for different values of $\Sigma$ and scaled with its value $\lambda_1^{(B,0)}$ without dissipation. 
 %
 By increasing the dissipation the absolute value of $|\lambda_1^{(B,0)}|/|\lambda_1^{(B)}|$ is enhanced, leading to a larger contribution to the prefactor. 

%
%
%
%
%
\begin{figure}[b!]
	\centering
	\includegraphics[width=0.8\linewidth]{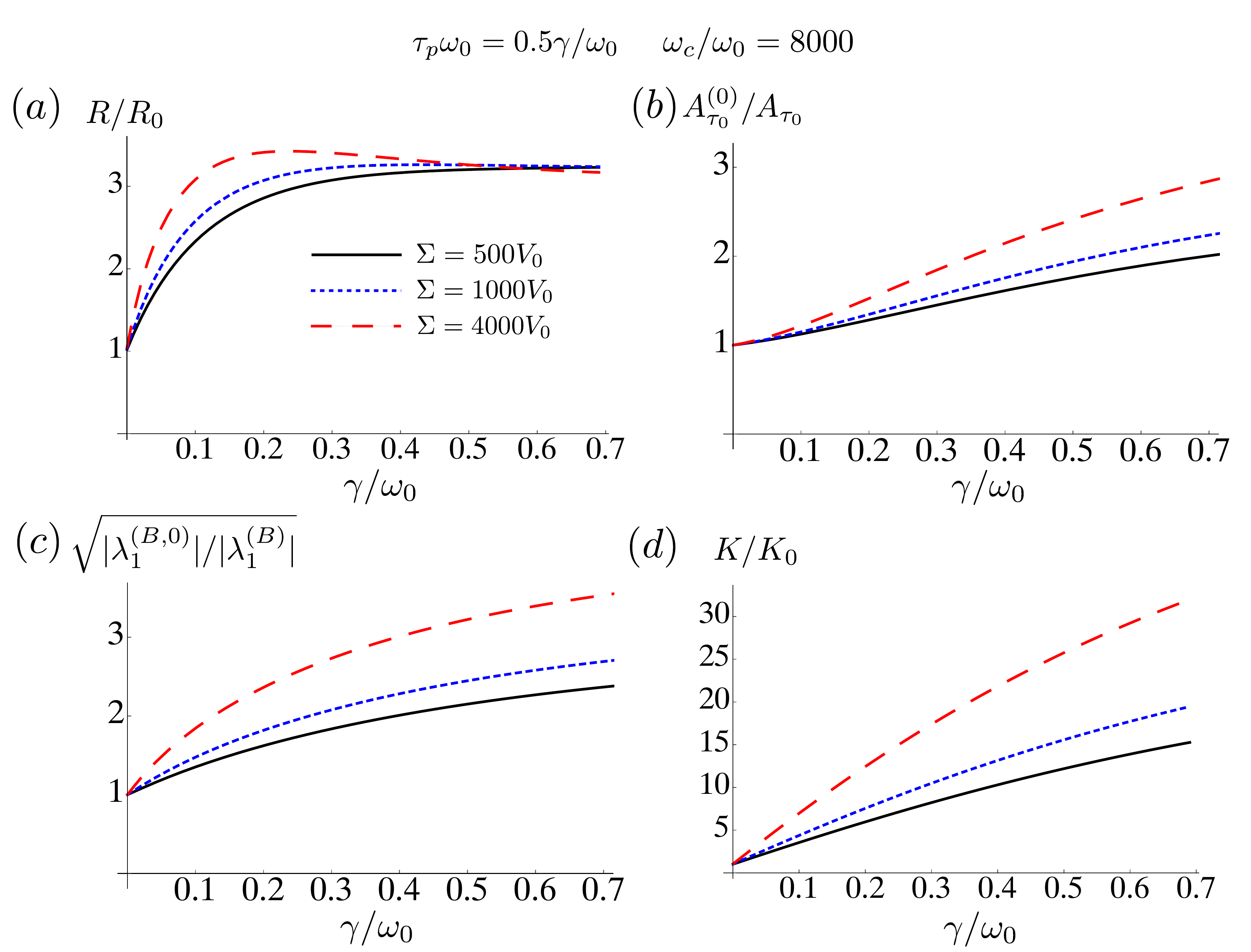}
	\caption{Change in the quantities contained in the prefactor $K$ due to dissipation for $\tau_p\omega_0=0.5 \gamma/\omega_0$ and $\omega_c/\omega_0=8000$. (a) Ratio of the determinants scaled with its value without dissipation, (b) prefactor of the Jacobian transformation scaled with  its value without dissipation, (c) nagative eigenvalue scaled with  its value without dissipation. (d) The prefactor $K$ of the escape rate scaled with its value $K_0$ without dissipation.  }
\label{FigK}
\end{figure}

\subsection{The Jacobian prefactor $A_{\tau_0}$}
Finally we discuss the result for the Jacobian prefactor  [\onlinecite{Maile2020,Grabert1987}]
\begin{align}
\frac{1}{A_{\tau_0}}=\sqrt{\int_{-\beta/2}^{{\beta/2}}d\tau \left(\dot{x}^{(1)}_{cl}(\tau)\right)^2}
\; .
\end{align}
In the non dissipative case we find the analytic expression
\begin{align}
\int_{-\beta/2}^{{\beta/2}}d\tau \left(\dot{x}^{(1)}_{cl}(\tau)\right)^2&\approx 
\frac{a^2\omega_{0}}{2}\left(\sqrt{1+\frac{\Sigma}{V_0}}+1\right)^{2} \left(1- e^{-\omega_0\xi_B^{(0)}}(1+\omega_0\xi^{(0)}_B)\right)\;.
\end{align}
The numerical value ${A_{\tau_0}}$ scaled with ${A_{\tau_0}^{(0)}}$ is shown in Fig.~\ref{FigK}b. 
%
%

\subsection{Calculation of the phases $\phi_{\lambda}^{\pm}$ }\label{subsec.phases}
In this section we give a detailed derivation of the quantities $\phi_{\lambda}^{\pm}$ appearing Eq.~(\ref{Eq.lnR}).

We start by calculating the Greens function $G_{\lambda}^{(0)}(\tau)$. 
In presence of Ohmic momentum dissipation without high frequency cutoff this quantity diverges similarly to the position quantum fluctuations.
We recall the dissipative kernel in Matsubara space for momentum dissipation with cutoff $\omega_c$   
\begin{equation}
 F^{(p)}(\omega)=\frac{-\tau_{p}|\omega|m f_c(\omega)}{1+\tau_{p}|\omega|f_c(\omega)}\;,
\end{equation}
with $f_c(\omega)=(1+|\omega|/\omega_c)^{-1}$. 
We rewrite  Eq.~(\ref{Eq.metagreendis}) and find 
\begin{align}
G_{\lambda}^{(0)}(\tau)\label{Eq. Fourierintegral}
&=\frac{1}{\pi m\omega_{c}}\int_{0}^{\infty}dx\frac{\left(1+(1+\tau_{p}\omega_{c})x\right)\cos(x\omega_{c}\tau)}{-p^{2}\Omega_{c}^{2}+\chi(-p^2) x+\alpha x^{2}+x^{3}-i\epsilon } 
, \quad (\epsilon \rightarrow 0)\;, 
\end{align}
with $p^2=-1+\lambda/m\omega_0^2$, 
$\Omega_c=\omega_0/\omega_c$, $\alpha={\gamma}/{\omega_{c}}+\tau_{p}\gamma+1$,
and
\begin{align}
\chi(-p^2)&=-p^{2}\Omega_{c}^{2}(1+\tau_{p}\omega_{c})+\frac{\gamma}{\omega_{c}}\;.
\end{align}
The denominator of Eq.~(\ref{Eq. Fourierintegral}) is a cubic polynomial with an imaginary part. 
We expand the polynomial into its roots $\tilde{x}_{1,2,3}$ and obtain, by introducing $\tilde{x}_{1}=\nu_1$, $\tilde{x}_{2}=-\nu_2$ and $\tilde{x}_{3}=-\nu_3$,
\begin{align}
-p^{2}\Omega_{c}^{2}+\chi(-p^2)& x+\alpha  \label{Eq.greenpol} x^{2}+x^{3}=(x-\nu_1)(x+\nu_2)(x+\nu_3)\;,
\end{align}
where $\nu_{1,2,3}>0$. 
Because the full form of the quantities $\nu_{1,2,3}$ is not important at this stage, we do not present them here explicitly, but refer to the next section. 
With this definition we can perform a principle value integration in Eq.~(\ref{Eq. Fourierintegral}). 
We obtain the result
\begin{align}
G_\lambda^{(0)}(\tau)
&=\frac{1}{\pi m\omega_{c}}\left(\sum_{i=1}^{3}\widetilde{\mathcal{U}}_ig[\tau\omega_c\nu_i]-\widetilde{\mathcal{U}}_1\pi\sin(\tau\omega_c\nu_1)\right)+\frac{i}{ m\omega_{c}}\frac{\left(1+(1+\tau_{p}\omega_{c})\nu_{1}\right)\cos(\nu_{1}\omega_{c}\tau)}{|\chi(-p^{2})+2\alpha\nu_{1}+3\nu_{1}^{2}|}\;,\label{Eq.finalgreentau}
\end{align}
where  the prefactors $\widetilde{\mathcal{U}}_i$ originate from a partial fraction expansion  (defined in the next section) 
and the auxiliary function $g(x)$ is defined in the previous section. 
For the factor $U^{-1}$ we have to calculate
\begin{align}
\dot{x}^{(2)}_{cl}\left(\frac{\xi_B}{2}\right)&=\frac{2\omega_{0}^{2}a_{0}}{\pi}\int_{0}^{\infty}dx\frac{\left(1+(1+\tau_{p}\omega_{c})x\right)\left[1-\cos(\xi_B\omega)\right]}{\Omega_{c}^{2}+\chi(1) x+\alpha x^{2}+x^{3}}\;, \label{Eq.U^-1}
\end{align}
which has no imaginary part. We calculate the integral by rewriting the polynomial in the denominator as
\begin{align}
\Omega_{c}^{2}+\chi(1) & x+\alpha  x^{2}+x^{3}=(x+k_1)(x+k_2)(x+k_3)\;, \label{Eq.upol}
\end{align}
with $\text{Re}(k_i)>0$, and find 
\begin{align}
U^{-1}=-\frac{1}{\pi m\omega_c}\sum_{i=1}^{3}\widetilde{\mathcal{T}}_i\left(\ln\left(k_i\right)+g[k_i\omega_c\xi]\right)\;,
\end{align}
where the prefactors $\widetilde{\mathcal{T}}_i$ and the quantities $k_i$ are also defined in the next section.
Inserting $\tau=\xi_B$ in Eq.~(\ref{Eq.finalgreentau}) and calculating $G_p^{(0)}(\tau=0)$ we obtain the result 
\begin{align}
n_\lambda^{(\pm)}=&-\frac{1}{\pi m\omega_c}\sum_{i=1}^{3}\widetilde{\mathcal{T}}_i\left(\ln\left(k_i\right)+g[k_i\omega_c\xi_B]\right)\nonumber+\frac{1}{\pi m\omega_{c}}\left(\sum_{i=1}^{3}\widetilde{\mathcal{U}}_i\left(\ln(\nu_{i})\pm g[\xi_B\omega_c\nu_i]\right)\mp\widetilde{\mathcal{U}}_1\pi\sin(\xi_B\omega_c\nu_1)\right)\nonumber
\\&-\frac{i}{ m\omega_{c}}\frac{\left(1+(1+\tau_{p}\omega_{c})\nu_{1}\right)\left(1\mp\cos(\nu_{1}\omega_{c}\xi_B)\right)}{|\chi(-p^{2})+2\alpha\nu_{1}+3\nu_{1}^{2}|}\;. \label{Eq.npm}
\end{align}
The phases are then calculated via $\phi_{\lambda}^{(\pm)}=\text{arg}(n_\lambda^{(\pm)})$ and we use $\phi_{m\omega_0^2}^{(\pm)}=-\pi$ and $\lim\limits_{\lambda\rightarrow \infty}\phi_{\lambda}^{(\pm)}=0$ in Eq.~(\ref{Eq.lnR}).

%
%
\subsection{Further auxiliary variables }
\label{Sec.Aux}
In the previous section we introduced the quantities  $\widetilde{\mathcal{T}}_i$, $\widetilde{\mathcal{U}}_i$, $\nu_i$, and $k_i$. 
The first two originate from the partial fraction expansions of the integrands in Eqs.~(\ref{Eq. Fourierintegral}) and (\ref{Eq.U^-1}).  The latter ones are related to the roots of the denominators of $\widetilde{G}^{(0)}_\lambda(\omega)$ and $U^{-1}$. The prefactors read
\begin{align}
\widetilde{\mathcal{T}}_1=\frac{1-(1+\tau_p \omega_c){k}_1}{({k}_1-{k}_2)({k}_1-{k}_3)} \quad\quad \widetilde{\mathcal{T}}_2=\frac{-1+(1+\tau_p \omega_c){k}_2}{({k}_1-{k}_2)({k}_2-{k}_3)}\quad\quad \widetilde{\mathcal{T}}_3=\frac{-1+(1+\tau_p \omega_c)k_3}{({k}_1-{k}_3)({k}_3-{k}_2)}
\end{align}
and
\begin{align}
\widetilde{\mathcal{U}}_1=\frac{1+(1+\tau_p \omega_c){\nu}_1}{(\nu_1+{\nu}_2)({\nu}_1+{\nu}_3)}\quad\quad \widetilde{\mathcal{U}}_2=\frac{1-(1+\tau_p \omega_c){\nu}_2}{({\nu}_1+{\nu}_2)({\nu}_2-{\nu}_3)}\quad\quad \widetilde{\mathcal{U}}_3=\frac{1-(1+\tau_p \omega_c)\nu_3}{(\nu_1+{\nu}_3)({\nu}_3-{\nu}_2)}\;.
\end{align}
Using the basic formula for the roots of cubic polynomials, we find for the ones of  Eq.~(\ref{Eq.upol})
\begin{align}
-k_1&=-\left(\frac{\alpha}{3}+\frac{2^{\frac{1}{3}}\eta(1)}{3\left(\Sigma(1)+\sqrt{4\eta^{3}(1)+\Sigma^{2}}\right)^{\frac{1}{3}}}-\frac{\left(\Sigma(1)+\sqrt{4\eta^{3}(1)+\Sigma^{2}(1)}\right)^{\frac{1}{3}}}{3\cdot2^{\frac{1}{3}}}\right)\nonumber\\
-k_2&=-\left(\frac{\alpha}{3}-\frac{(1+i\sqrt{3})\eta(1)}{3\cdot2^{\frac{2}{3}}\left(\Sigma(1)+\sqrt{4\eta^{3}(1)+\Sigma^{2}(1)}\right)^{\frac{1}{3}}}+\frac{(1-i\sqrt{3})\left(\Sigma(1)+\sqrt{4\eta(1)+\Sigma^{2}(1)}\right)^{\frac{1}{3}}}{6\cdot2^{\frac{1}{3}}}\right)\\
-k_3&=-\left(\frac{\alpha}{3}-\frac{(1-i\sqrt{3})\eta(1)}{3\cdot2^{\frac{2}{3}}\left(\Sigma(1)+\sqrt{4\eta^{3}(1)+\Sigma^{2}(1)}\right)^{\frac{1}{3}}}+\frac{(1+i\sqrt{3})\left(\Sigma(1)+\sqrt{4\eta(1)+\Sigma^{2}(1)}\right)^{\frac{1}{3}}}{6\cdot2^{\frac{1}{3}}}\right)\nonumber\;,
\end{align}
where we defined $\eta(1)=3\chi(1)-\alpha^{2}$ and $\Sigma(1)=9\alpha\chi(1)-2\alpha^{3}-27\Omega_{c}^{2}$ 
(and used 
$p^2=-1+\lambda/m\omega_0^2$, $\Omega_c=\omega_0/\omega_c$, $\alpha={\gamma}/{\omega_{c}}+\tau_{p}\gamma+1$).
Further, the roots for the polynomial (\ref{Eq.greenpol}) read
\begin{align}
\nu_1&=-\left(\frac{\alpha}{3}+\frac{2^{\frac{1}{3}}\eta(p^2)}{3\left(\Sigma(p^{2})+\sqrt{4\eta^{3}(p^{2})+\Sigma^{2}(p^{2})}\right)^{\frac{1}{3}}}-\frac{\left(\Sigma(p^{2})+\sqrt{4\eta^{3}(p^{2})+\Sigma^{2}(p^{2})}\right)^{\frac{1}{3}}}{3\cdot2^{\frac{1}{3}}}\right)\nonumber\\
-\nu_2&=-\left(\frac{\alpha}{3}-\frac{(1+i\sqrt{3})\eta(p^{2})}{3\cdot2^{\frac{2}{3}}\left(\Sigma(p^{2})+\sqrt{4\eta^{3}(p^{2})+\Sigma^{2}(p^{2})}\right)^{\frac{1}{3}}}+\frac{(1-i\sqrt{3})\left(\Sigma(p^{2})+\sqrt{4\eta^3(p^{2})+\Sigma^{2}(p^{2})}\right)^{\frac{1}{3}}}{6\cdot2^{\frac{1}{3}}}\right)\\
-\nu_3&=-\left(\frac{\alpha}{3}-\frac{(1-i\sqrt{3})\eta(p^2)}{3\cdot2^{\frac{2}{3}}\left(\Sigma(p^{2})+\sqrt{4\eta^{3}(p^{2})+\Sigma^{2}(p^{2})}\right)^{\frac{1}{3}}}+\frac{(1+i\sqrt{3})\left(\Sigma(p^{2})+\sqrt{4\eta^3(p^{2})+\Sigma^{2}(p^{2})}\right)^{\frac{1}{3}}}{6\cdot2^{\frac{1}{3}}}\right)\nonumber\;,
\end{align}
where $\eta(p^2)=3\chi(-p^{2})-\alpha^{2}$ and
$\Sigma(p^2)=9\alpha\chi(-p^{2})-2\alpha^{3}+27p^{2}\Omega_{c}^{2}$.

%
%
%
%

\section{The tunneling average energy loss in presence of dissipation}\label{App:D}
\begin{figure}[t]
	\centering
	\includegraphics[width=0.8\linewidth]{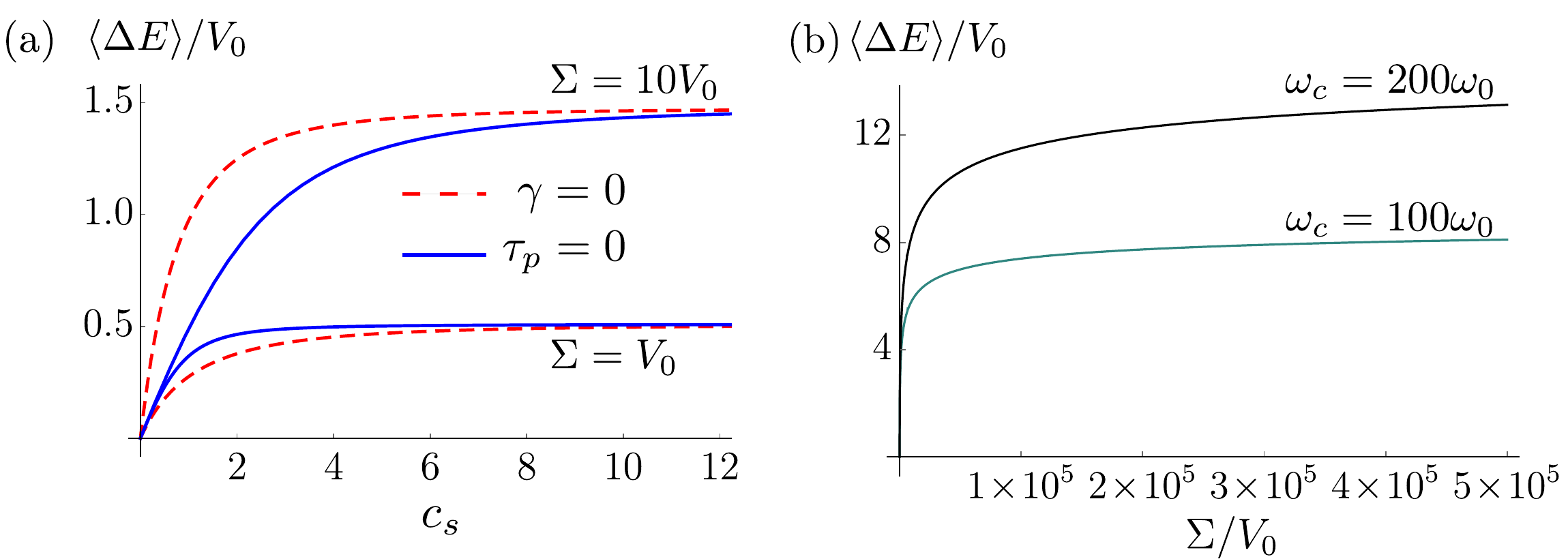}
	\caption{(a) The average energy loss $\langle \Delta E \rangle$ as a function of the dissipative coupling strength, $c_s = \gamma/\omega_0$ for position dissipation and $c_s=\omega_0 \tau_p$, for two values of $\Sigma$, 
	for pure position dissipation (solid blue line) and for pure momentum dissipation (dashed red line)
	In the overdamped limit  $\langle \Delta E \rangle$ saturate to the same value.  
	(b) Saturation of the energy loss in presence with pure momentum dissipation as a function of $\Sigma$, for different $\omega_c$ at $\tau_p\omega_0=0.5$.}
	\label{Fig.Fig3-supp}
\end{figure}
 As discussed in the main text the returning point can be used to calculate
the average energy loss  $\langle \Delta E \rangle $ of the particle during the tunneling  in the presence of the dissipative interaction with the environment according the equation
$
\langle \Delta E \rangle =V(0)-V(x_{esc})
$
\cite{Grabert1987,Weiss1984}.

In Fig.~\ref{Fig.Fig3-supp}a we show the results for $\langle \Delta E \rangle$ as a function of the dissipative coupling strength 
defined as $c_S=\gamma/\omega_0$ for the position dissipation and as $c_S=\omega_0 \tau_p$ for the momentum dissipation.

In Fig.~\ref{Fig.Fig3-supp}a the blue solid line corresponds to pure position dissipation ($c_S=\gamma/\omega_0$, $\tau_p=0$) 
while the red dashed line to the results for pure momentum dissipation  ($c_S=\tau_p\omega_0$, $\gamma =0$).
In the overdamped limit $c_S \gg 1$ both dissipative environments saturate to the same energy loss, at fixed value of $\Sigma$.
For small dissipative couplings, at $\Sigma=V_0$ position dissipation dissipates more energy than momentum dissipation whereas 
for  $\Sigma=10V_0$ the situation is reversed: momentum dissipation has a larger influence. 
 
In Fig.~\ref{Fig.Fig3-supp}b we fix the dissipative coupling strength and show $\langle \Delta E\rangle$ as a function of $\Sigma/V_0$ for pure momentum dissipation. 
The loss saturates, similarly to the escape rate,  to a value determined by the high frequency cutoff $\omega_c$.

\section{Perturbative results for smooth metastable potentials}\label{App:E}
In this section, we show that the enhancement of the escape rate via momentum dissipation, discussed for the semi-double parabolic potential in the main text, is also valid for more general metastable potentials. We focus the discussion on pure momentum dissipation and use a perturbative approach to find an approximate action. 
In the analysis presented here, as in the main text, we  neglect the influence of the prefactor on the decay rate as the exponential part containing the action is the leading term.\\

Specifically, we consider a potential having a smooth barrier top, by inverting the parabola on the right side 
\begin{align}
    V(x)=\frac{m}{2}\begin{cases}
    \omega_{0}^{2}(x+a)^{2} &\;\;\;\; \text{for}\;\;\;\; x<0 \\ -\omega_{B}^{2}\left[(x-c\cdot a)^{2}-2d\right] &\;\;\;\; \text{for}\;\;\;\; x>0
    \end{cases},\label{Eq.newpot}
\end{align}
where $c=\omega_{0}^{2}/\omega_{B}^{2}$ and $d=\frac{a^{2}c}{2}\left(1+c\right)$.  $\omega_0$ is the frequency of the well-parabola and $\omega_B$ the frequency of barrier-parabola.
We show the potential in Fig.~\ref{Fig.Fig5-supp}a: the potential and its derivative are continuous functions at the point $x=0$.
In the non dissipative case the bounce path satisfies the differential equation
\begin{align}
-\ddot{x}_{cl}(\tau)+\omega_{0}^{2}\Theta(-x_{cl}(\tau))x_{cl}(\tau)-\omega_{B}^{2}\Theta(x_{cl}(\tau))x_{cl}(\tau)&=-\omega_{0}^{2}a
\end{align}
where $-a$ is the position of the minimum of the left parabola and we find the bounce solution
\begin{align}
    x_{cl}(\tau)=\Theta(-(\tau\!+\!\tau_{1}))\left(-a\!+\!A(\tau_{1})e^{\omega_{0}(\tau+\tau_{1})}\right)+\Theta(\tau_{1}\!-\!|\tau|)\left(ca\!+\!B(\tau_{1})\cos(\omega_{B}\tau)\right)+\Theta(\tau\!-\!\tau_{1})\left(-a\!+\!A(\tau_{1})e^{-\omega_{0}(\tau-\tau_{1})}\right), \label{Eq.newbounce}
\end{align}
with $A(\tau_1)=(c+1)a+B(\tau_1)\cos(\omega_{B}\tau_{1})$ and 
\begin{align}
B(\tau_1)=\frac{\omega_{0}(c+1)a}{\left(\omega_{B}\sin(\omega_{B}\tau_{1})-\omega_{0}\cos(\omega_{B}\tau_{1})\right)}.
\end{align}
Here, $-\tau_1$ is the point in imaginary time at which the path crosses from the left to the right parabola and $\tau_1$ is the point of the opposite event, namely 
$x(-\tau_1)=x(\tau_1)=0$.
In the non dissipative regime and in the symmetric case ($\omega_B/\omega_0=1$), we find  $\tau_1\omega_0=3/4\pi$.
\begin{figure}[t]
	\centering
	\includegraphics[width=0.8\linewidth]{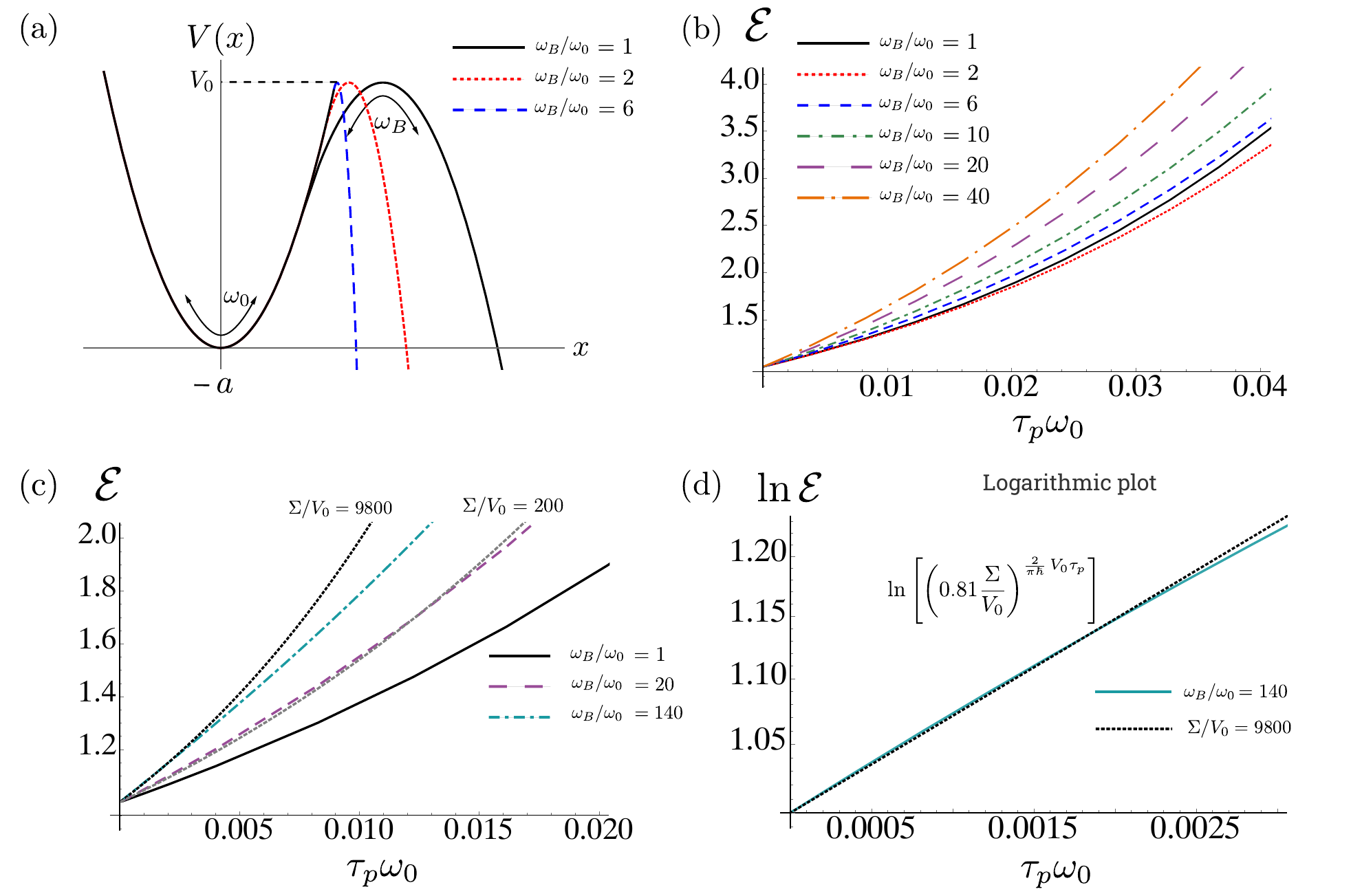}
	\caption{(a) 
	Smooth potential studied in this section. By increasing $\omega_B/\omega_0$ the potential becomes comparable to the semi-double parabolic  potential. (b) Perturbative results for $\mathcal{E}$ for different ratios $\omega_B/\omega_0$ as a function of $\tau_p\omega_0$.   (c) Comparison between the results for the semi-double parabolic and for the smooth potential as discussed in the text. Dotted lines correspond to results for the semi-double parabolic potential while coloured lines are for the smooth potential. (d) Logarithmic plot of the analytical formula Eq.(8) in the main text and the perturbative results for the smooth potential. Both potentials have the same slope at the turning point $dV(x)/dx|_{x_{esc}}=-700 m\omega_0^2$. }
	\label{Fig.Fig5-supp}
\end{figure}
Increasing $\omega_B$ the barrier thickness decreases and for large ratios $\omega_B/\omega_0\gg 1$ 
the barrier becomes comparable to the semi-double parabolic potential of the main text in the regime $\Sigma/V_0 \gg 1$ (sharp potential), 
but with a smooth barrier top.
Note that $2\tau_1$ plays the role of the bounce time in the latter case.\\

To qualitatively compare the influence of momentum dissipation on both potentials, we choose the limit of low dissipative couplings $\tau_p\omega_0\ll 1$.
In this regime, we assume that the bounce for the potential Eq.~(\ref{Eq.newpot}), which is parametrized by Eq.~(\ref{Eq.newbounce}), 
does not significantly change due to the dissipation.
Hence, we insert the non-dissipatve bounce Eq.~(\ref{Eq.newbounce}) into the full action containing the momentum dissipation.
In the following results, we fix  the barrier height to $V_0/(\hbar\omega_0)=12.5$ and consider  different frequency ratios $\omega_B/\omega_0$. Note that in this way the position of the minimum changes as 
$(-a)=-\sqrt{\frac{V_{0}}{1+\omega_{0}^{2} / \omega_{B}^{2}}}$.\\

Fig.~\ref{Fig.Fig5-supp}b shows the enhancement   $\mathcal{E}$ for different ratios of $\omega_B/\omega_0$ as a function of $\tau_p\omega_0$.
We remark that  the enhancement for the escape still occurs.
The influence of momentum dissipation increases by increasing $\omega_B/\omega_0$, as expected from the analysis given in the main text, namely 
the derivative of the potential at the turning point increases by increasing  $\omega_B/\omega_0$. 
However, as shown by the red and the black line in Fig.~\ref{Fig.Fig5-supp}b, for $\omega_B/\omega_0 =1$ and $\omega_B/\omega_0 =2$, 
the trend of the influence of momentum dissipation is reversed.
The enhancement for $\omega_B/\omega_0 =1$ is stronger than for $\omega_B/\omega_0 =2$, although the latter potential has a larger slope at the turning point.
This can be explained by the fact that by varying the ratio one also modifies significantly the shape of the potential leading to a different parametric prefactor in the action which is important in the weak dissipative coupling regime.
This is an accidental effect which disappears as long as we consider large ratio $\omega_B/\omega_0 \geq 2$.\\

In Fig.~\ref{Fig.Fig5-supp}c, we plot the cases $\omega_B/\omega_0=1$, $\omega_B/\omega_0=20$ and $\omega_B/\omega_0=140$ to compare the results for the smooth potential 
Eq.~(\ref{Eq.newpot}) with the semi-double parabolic potential of the main text.
We make a connection between the results for the two different potentials by looking at their slope at the turning  $x_{esc}$. 
For $\omega_B/\omega_0=20$, we find $dV(x)/dx|_{x_{esc}}=-100 m\omega_0^2$, which corresponds to $\Sigma/V_0=200$ for the semi-double parabolic  potential. 
Both results agree well for the regime displayed. 
In the case of $\omega_B/\omega_0=140$, the slope $dV(x)/dx|_{x_{esc}}=-700 m\omega_0^2$ corresponding to $\Sigma/V_0=9800$ 
for the semi-double parabolic  potential. 
Here, we find a good agreement of both result for values below $\tau_p\omega_0\approx 0.0025$.\\

The deviations between the two different potentials appear  beyond some value of $\tau_p\omega_0$,  which depends on the ratio $\omega_B/\omega_0$.
For example, we have a good agreement until $\tau_p\omega_0=0.02$  for $\omega_B/\omega_0=20$ and
until $\tau_p\omega_0\approx 0.0025$ for $\omega_B/\omega_0=140$.
In Fig.~\ref{Fig.Fig5-supp}d, we plot the log scale the case  $\omega_B/\omega_0=140$.
This deviation can be explained by the break down of the non-dissipative bounce approximation.
Indeed, the influence of momentum dissipation on the bounce path is expected to be relevant when the bounce becomes narrower, e.g. when $\omega_B/\omega_0$ is increased. 
As explained in the main text, momentum dissipation additionally squeezes the bounce.
Hence, by inserting the non dissipative bounce, we are underestimating the effect of the momentum dissipation.
For this reason, the results for the smooth potential deviates from the ones of the semi-double parabolic  potential, in 
which we take the dissipative interaction for calculating the bounce path into account.\\

To summarize, the results presented in the main text obtained for the semi-double parabolic potential - which has a singular behavior of the derivative at the top barrier - 
are valid for the general case as we proved that similar results hold for a potential with a smooth barrier top. 
We showed the comparison only in the weak dissipative interaction limit for which we have an approximated solution for the potential with a smooth barrier top.  
In Fig.~\ref{Fig.Fig5-supp}d, we directly compare the analytical result of Eq.~(8) of the main text, which are valid in the limit $\Sigma/V_0 \gg 1$ and $\tau_p\omega_0 \ll 1$ (but $\xi_B\omega_c\ll1$), with the results achievable for the smooth potential for the case $\omega_B/\omega_0 \gg 1$.
The results almost coincides for values up to $\tau_p\omega_0\approx 0.0025$ and then starts to deviate because of the above explained reasons.
This points out that, at least in the small coupling limit, the results for the smooth potential are the same as for the semi-double parabolic  potential. 
We also find that the enhancement for the potential  with a smooth barrier top can be even larger than the 
enhancement for the semi-double parabolic  potential in some parameters range.

\end{widetext}

\bibliographystyle{mybibstyle}
 
\bibliography{bibliographymeta}

\begin{thebibliography}{45}%
\makeatletter
\providecommand \@ifxundefined [1]{%
 \@ifx{#1\undefined}
}%
\providecommand \@ifnum [1]{%
 \ifnum #1\expandafter \@firstoftwo
 \else \expandafter \@secondoftwo
 \fi
}%
\providecommand \@ifx [1]{%
 \ifx #1\expandafter \@firstoftwo
 \else \expandafter \@secondoftwo
 \fi
}%
\providecommand \natexlab [1]{#1}%
\providecommand \emph  [1]{``#1''}%
\providecommand \bibnamefont  [1]{#1}%
\providecommand \bibfnamefont [1]{#1}%
\providecommand \citenamefont [1]{#1}%
\providecommand \href@noop [0]{\@secondoftwo}%
\providecommand \href [0]{\begingroup \@sanitize@url \@href}%
\providecommand \@href[1]{\@@startlink{#1}\@@href}%
\providecommand \@@href[1]{\endgroup#1\@@endlink}%
\providecommand \@sanitize@url [0]{\catcode `\\12\catcode `\$12\catcode
  `\&12\catcode `\#12\catcode `\^12\catcode `\_12\catcode `\%12\relax}%
\providecommand \@@startlink[1]{}%
\providecommand \@@endlink[0]{}%
\providecommand \url  [0]{\begingroup\@sanitize@url \@url }%
\providecommand \@url [1]{\endgroup\@href {#1}{\urlprefix }}%
\providecommand \urlprefix  [0]{URL }%
\providecommand \Eprint [0]{\href }%
\providecommand \doibase [0]{http://dx.doi.org/}%
\providecommand \selectlanguage [0]{\@gobble}%
\providecommand \bibinfo  [0]{\@secondoftwo}%
\providecommand \bibfield  [0]{\@secondoftwo}%
\providecommand \translation [1]{[#1]}%
\providecommand \BibitemOpen [0]{}%
\providecommand \bibitemStop [0]{}%
\providecommand \bibitemNoStop [0]{.\EOS\space}%
\providecommand \EOS [0]{\spacefactor3000\relax}%
\providecommand \BibitemShut  [1]{\csname bibitem#1\endcsname}%
\let\auto@bib@innerbib\@empty
\bibitem [{\citenamefont {Santoro}\ and\ \citenamefont
  {Tosatti}(2006)}]{Tosatti2006}%
  \BibitemOpen
  \bibfield  {author} {\bibinfo {author} {\bibfnamefont {G.~E.}\ \bibnamefont
  {Santoro}}\ and\ \bibinfo {author} {\bibfnamefont {E.}~\bibnamefont
  {Tosatti}},\ }\bibfield  {title} {\emph {\bibinfo {title} {Optimization using
  quantum mechanics: quantum annealing through adiabatic evolution},}\ }\href
  {\doibase 10.1088/0305-4470/39/36/r01} {\bibfield  {journal} {\bibinfo
  {journal} {Journal of Physics A: Mathematical and General}\ }\textbf
  {\bibinfo {volume} {39}},\ \bibinfo {pages} {R393} (\bibinfo {year} {2006})}\
  \BibitemShut {NoStop}%
\bibitem [{\citenamefont {Stella}\ \emph {et~al.}(2005)\citenamefont {Stella},
  \citenamefont {Santoro},\ and\ \citenamefont {Tosatti}}]{Stella2005}%
  \BibitemOpen
  \bibfield  {author} {\bibinfo {author} {\bibfnamefont {L.}~\bibnamefont
  {Stella}}, \bibinfo {author} {\bibfnamefont {G.~E.}\ \bibnamefont {Santoro}},
  \ and\ \bibinfo {author} {\bibfnamefont {E.}~\bibnamefont {Tosatti}},\
  }\bibfield  {title} {\emph {\bibinfo {title} {Optimization by quantum
  annealing: Lessons from simple cases},}\ }\href {\doibase
  10.1103/PhysRevB.72.014303} {\bibfield  {journal} {\bibinfo  {journal} {Phys.
  Rev. B}\ }\textbf {\bibinfo {volume} {72}},\ \bibinfo {pages} {014303}
  (\bibinfo {year} {2005})}\ \BibitemShut {NoStop}%
\bibitem [{\citenamefont {Albash}\ and\ \citenamefont
  {Lidar}(2018)}]{Lidar20182}%
  \BibitemOpen
  \bibfield  {author} {\bibinfo {author} {\bibfnamefont {T.}~\bibnamefont
  {Albash}}\ and\ \bibinfo {author} {\bibfnamefont {D.~A.}\ \bibnamefont
  {Lidar}},\ }\bibfield  {title} {\emph {\bibinfo {title} {Adiabatic quantum
  computation},}\ }\href {\doibase 10.1103/RevModPhys.90.015002} {\bibfield
  {journal} {\bibinfo  {journal} {Rev. Mod. Phys.}\ }\textbf {\bibinfo {volume}
  {90}},\ \bibinfo {pages} {015002} (\bibinfo {year} {2018})}\ \BibitemShut
  {NoStop}%
\bibitem [{\citenamefont {Hauke}\ \emph {et~al.}(2020)\citenamefont {Hauke},
  \citenamefont {Katzgraber}, \citenamefont {Lechner}, \citenamefont
  {Nishimori},\ and\ \citenamefont {Oliver}}]{Hauke2020}%
  \BibitemOpen
  \bibfield  {author} {\bibinfo {author} {\bibfnamefont {P.}~\bibnamefont
  {Hauke}}, \bibinfo {author} {\bibfnamefont {H.~G.}\ \bibnamefont
  {Katzgraber}}, \bibinfo {author} {\bibfnamefont {W.}~\bibnamefont {Lechner}},
  \bibinfo {author} {\bibfnamefont {H.}~\bibnamefont {Nishimori}}, \ and\
  \bibinfo {author} {\bibfnamefont {W.~D.}\ \bibnamefont {Oliver}},\ }\bibfield
   {title} {\emph {\bibinfo {title} {Perspectives of quantum annealing: methods
  and implementations},}\ }\href {\doibase 10.1088/1361-6633/ab85b8} {\bibfield
   {journal} {\bibinfo  {journal} {Reports on Progress in Physics}\ }\textbf
  {\bibinfo {volume} {83}},\ \bibinfo {pages} {054401} (\bibinfo {year}
  {2020})}\ \BibitemShut {NoStop}%
\bibitem [{\citenamefont {Durkin}(2019)}]{Durkin2019}%
  \BibitemOpen
  \bibfield  {author} {\bibinfo {author} {\bibfnamefont {G.~A.}\ \bibnamefont
  {Durkin}},\ }\bibfield  {title} {\emph {\bibinfo {title} {Quantum speedup at
  zero temperature via coherent catalysis},}\ }\href {\doibase
  10.1103/PhysRevA.99.032315} {\bibfield  {journal} {\bibinfo  {journal} {Phys.
  Rev. A}\ }\textbf {\bibinfo {volume} {99}},\ \bibinfo {pages} {032315}
  (\bibinfo {year} {2019})}\ \BibitemShut {NoStop}%
\bibitem [{\citenamefont {Denchev}\ \emph {et~al.}(2016)\citenamefont
  {Denchev}, \citenamefont {Boixo}, \citenamefont {Isakov}, \citenamefont
  {Ding}, \citenamefont {Babbush}, \citenamefont {Smelyanskiy}, \citenamefont
  {Martinis},\ and\ \citenamefont {Neven}}]{Neven2016}%
  \BibitemOpen
  \bibfield  {author} {\bibinfo {author} {\bibfnamefont {V.~S.}\ \bibnamefont
  {Denchev}}, \bibinfo {author} {\bibfnamefont {S.}~\bibnamefont {Boixo}},
  \bibinfo {author} {\bibfnamefont {S.~V.}\ \bibnamefont {Isakov}}, \bibinfo
  {author} {\bibfnamefont {N.}~\bibnamefont {Ding}}, \bibinfo {author}
  {\bibfnamefont {R.}~\bibnamefont {Babbush}}, \bibinfo {author} {\bibfnamefont
  {V.}~\bibnamefont {Smelyanskiy}}, \bibinfo {author} {\bibfnamefont
  {J.}~\bibnamefont {Martinis}}, \ and\ \bibinfo {author} {\bibfnamefont
  {H.}~\bibnamefont {Neven}},\ }\bibfield  {title} {\emph {\bibinfo {title}
  {What is the Computational Value of Finite-Range Tunneling?}}\ }\href
  {\doibase 10.1103/PhysRevX.6.031015} {\bibfield  {journal} {\bibinfo
  {journal} {Phys. Rev. X}\ }\textbf {\bibinfo {volume} {6}},\ \bibinfo {pages}
  {031015} (\bibinfo {year} {2016})}\ \BibitemShut {NoStop}%
\bibitem [{\citenamefont {Albash}\ and\ \citenamefont
  {Lidar}(2015)}]{Lidar2015}%
  \BibitemOpen
  \bibfield  {author} {\bibinfo {author} {\bibfnamefont {T.}~\bibnamefont
  {Albash}}\ and\ \bibinfo {author} {\bibfnamefont {D.~A.}\ \bibnamefont
  {Lidar}},\ }\bibfield  {title} {\emph {\bibinfo {title} {Decoherence in
  adiabatic quantum computation},}\ }\href {\doibase
  10.1103/PhysRevA.91.062320} {\bibfield  {journal} {\bibinfo  {journal} {Phys.
  Rev. A}\ }\textbf {\bibinfo {volume} {91}},\ \bibinfo {pages} {062320}
  (\bibinfo {year} {2015})}\ \BibitemShut {NoStop}%
\bibitem [{\citenamefont {Albash}\ \emph {et~al.}(2017)\citenamefont {Albash},
  \citenamefont {Martin-Mayor},\ and\ \citenamefont {Hen}}]{Hen2017}%
  \BibitemOpen
  \bibfield  {author} {\bibinfo {author} {\bibfnamefont {T.}~\bibnamefont
  {Albash}}, \bibinfo {author} {\bibfnamefont {V.}~\bibnamefont
  {Martin-Mayor}}, \ and\ \bibinfo {author} {\bibfnamefont {I.}~\bibnamefont
  {Hen}},\ }\bibfield  {title} {\emph {\bibinfo {title} {Temperature Scaling
  Law for Quantum Annealing Optimizers},}\ }\href {\doibase
  10.1103/PhysRevLett.119.110502} {\bibfield  {journal} {\bibinfo  {journal}
  {Phys. Rev. Lett.}\ }\textbf {\bibinfo {volume} {119}},\ \bibinfo {pages}
  {110502} (\bibinfo {year} {2017})}\ \BibitemShut {NoStop}%
\bibitem [{\citenamefont {Mishra}\ \emph {et~al.}(2018)\citenamefont {Mishra},
  \citenamefont {Albash},\ and\ \citenamefont {Lidar}}]{Lidar2018}%
  \BibitemOpen
  \bibfield  {author} {\bibinfo {author} {\bibfnamefont {A.}~\bibnamefont
  {Mishra}}, \bibinfo {author} {\bibfnamefont {T.}~\bibnamefont {Albash}}, \
  and\ \bibinfo {author} {\bibfnamefont {D.~A.}\ \bibnamefont {Lidar}},\
  }\bibfield  {title} {\emph {\bibinfo {title} {Finite temperature quantum
  annealing solving exponentially small gap problem with non-monotonic success
  probability},}\ }\href {\doibase 10.1038/s41467-018-05239-9} {\bibfield
  {journal} {\bibinfo  {journal} {Nature Communications}\ }\textbf {\bibinfo
  {volume} {9}},\ \bibinfo {pages} {2917} (\bibinfo {year} {2018})}\
  \BibitemShut {NoStop}%
\bibitem [{\citenamefont {Verstraete}\ \emph {et~al.}(2009)\citenamefont
  {Verstraete}, \citenamefont {Wolf},\ and\ \citenamefont
  {Ignacio~Cirac}}]{Verstraete2009}%
  \BibitemOpen
  \bibfield  {author} {\bibinfo {author} {\bibfnamefont {F.}~\bibnamefont
  {Verstraete}}, \bibinfo {author} {\bibfnamefont {M.}~\bibnamefont {Wolf}}, \
  and\ \bibinfo {author} {\bibfnamefont {J.}~\bibnamefont {Ignacio~Cirac}},\
  }\bibfield  {title} {\emph {\bibinfo {title} {Quantum computation and
  quantum-state engineering driven by dissipation},}\ }\href {\doibase
  https://doi.org/10.1038/nphys1342} {\bibfield  {journal} {\bibinfo  {journal}
  {Nature Phys.}\ }\textbf {\bibinfo {volume} {5}},\ \bibinfo {pages} {633}
  (\bibinfo {year} {2009})}\ \BibitemShut {NoStop}%
\bibitem [{\citenamefont {Plenio}\ and\ \citenamefont
  {Huelga}(2008)}]{Plenio_2008}%
  \BibitemOpen
  \bibfield  {author} {\bibinfo {author} {\bibfnamefont {M.~B.}\ \bibnamefont
  {Plenio}}\ and\ \bibinfo {author} {\bibfnamefont {S.~F.}\ \bibnamefont
  {Huelga}},\ }\bibfield  {title} {\emph {\bibinfo {title} {Dephasing-assisted
  transport: quantum networks and biomolecules},}\ }\href {\doibase
  10.1088/1367-2630/10/11/113019} {\bibfield  {journal} {\bibinfo  {journal}
  {New Journal of Physics}\ }\textbf {\bibinfo {volume} {10}},\ \bibinfo
  {pages} {113019} (\bibinfo {year} {2008})}\ \BibitemShut {NoStop}%
\bibitem [{\citenamefont {Mohseni}\ \emph {et~al.}(2008)\citenamefont
  {Mohseni}, \citenamefont {Rebentrost}, \citenamefont {Lloyd},\ and\
  \citenamefont {Aspuru-Guzik}}]{Mohseni2008}%
  \BibitemOpen
  \bibfield  {author} {\bibinfo {author} {\bibfnamefont {M.}~\bibnamefont
  {Mohseni}}, \bibinfo {author} {\bibfnamefont {P.}~\bibnamefont {Rebentrost}},
  \bibinfo {author} {\bibfnamefont {S.}~\bibnamefont {Lloyd}}, \ and\ \bibinfo
  {author} {\bibfnamefont {A.}~\bibnamefont {Aspuru-Guzik}},\ }\bibfield
  {title} {\emph {\bibinfo {title} {Environment-assisted quantum walks in
  photosynthetic energy transfer},}\ }\href {\doibase
  https://doi.org/10.1063/1.3002335} {\bibfield  {journal} {\bibinfo  {journal}
  {J. Chem. Phys.}\ }\textbf {\bibinfo {volume} {129}},\ \bibinfo {pages}
  {174106} (\bibinfo {year} {2008})}\ \BibitemShut {NoStop}%
\bibitem [{\citenamefont {Maier}\ \emph {et~al.}(2019)\citenamefont {Maier},
  \citenamefont {Brydges}, \citenamefont {Jurcevic}, \citenamefont {Trautmann},
  \citenamefont {Hempel}, \citenamefont {Lanyon}, \citenamefont {Hauke},
  \citenamefont {Blatt},\ and\ \citenamefont {Roos}}]{Maier2019}%
  \BibitemOpen
  \bibfield  {author} {\bibinfo {author} {\bibfnamefont {C.}~\bibnamefont
  {Maier}}, \bibinfo {author} {\bibfnamefont {T.}~\bibnamefont {Brydges}},
  \bibinfo {author} {\bibfnamefont {P.}~\bibnamefont {Jurcevic}}, \bibinfo
  {author} {\bibfnamefont {N.}~\bibnamefont {Trautmann}}, \bibinfo {author}
  {\bibfnamefont {C.}~\bibnamefont {Hempel}}, \bibinfo {author} {\bibfnamefont
  {B.~P.}\ \bibnamefont {Lanyon}}, \bibinfo {author} {\bibfnamefont
  {P.}~\bibnamefont {Hauke}}, \bibinfo {author} {\bibfnamefont
  {R.}~\bibnamefont {Blatt}}, \ and\ \bibinfo {author} {\bibfnamefont {C.~F.}\
  \bibnamefont {Roos}},\ }\bibfield  {title} {\emph {\bibinfo {title}
  {Environment-Assisted Quantum Transport in a 10-qubit Network},}\ }\href
  {\doibase 10.1103/PhysRevLett.122.050501} {\bibfield  {journal} {\bibinfo
  {journal} {Phys. Rev. Lett.}\ }\textbf {\bibinfo {volume} {122}},\ \bibinfo
  {pages} {050501} (\bibinfo {year} {2019})}\ \BibitemShut {NoStop}%
\bibitem [{\citenamefont {Gamow}(1928)}]{Gamow1928}%
  \BibitemOpen
  \bibfield  {author} {\bibinfo {author} {\bibfnamefont {G.}~\bibnamefont
  {Gamow}},\ }\bibfield  {title} {\emph {\bibinfo {title} {Zur Quantentheorie
  des Atomkernes},}\ }\href {\doibase 10.1007/BF01343196} {\bibfield  {journal}
  {\bibinfo  {journal} {Zeitschrift f{\"u}r Physik}\ }\textbf {\bibinfo
  {volume} {51}},\ \bibinfo {pages} {204} (\bibinfo {year} {1928})}\
  \BibitemShut {NoStop}%
\bibitem [{\citenamefont {Sethna}(1982)}]{Sethnaimp1982}%
  \BibitemOpen
  \bibfield  {author} {\bibinfo {author} {\bibfnamefont {J.~P.}\ \bibnamefont
  {Sethna}},\ }\bibfield  {title} {\emph {\bibinfo {title} {Decay rates of
  tunneling centers coupled to phonons: An instanton approach},}\ }\href
  {\doibase 10.1103/PhysRevB.25.5050} {\bibfield  {journal} {\bibinfo
  {journal} {Phys. Rev. B}\ }\textbf {\bibinfo {volume} {25}},\ \bibinfo
  {pages} {5050} (\bibinfo {year} {1982})}\ \BibitemShut {NoStop}%
\bibitem [{\citenamefont {Callan}\ and\ \citenamefont
  {Coleman}(1977)}]{Callan1977}%
  \BibitemOpen
  \bibfield  {author} {\bibinfo {author} {\bibfnamefont {C.~G.}\ \bibnamefont
  {Callan}}\ and\ \bibinfo {author} {\bibfnamefont {S.}~\bibnamefont
  {Coleman}},\ }\bibfield  {title} {\emph {\bibinfo {title} {Fate of the false
  vacuum. II. First quantum corrections},}\ }\href {\doibase
  10.1103/PhysRevD.16.1762} {\bibfield  {journal} {\bibinfo  {journal} {Phys.
  Rev. D}\ }\textbf {\bibinfo {volume} {16}},\ \bibinfo {pages} {1762}
  (\bibinfo {year} {1977})}\ \BibitemShut {NoStop}%
\bibitem [{\citenamefont {Coleman}(1979)}]{Coleman1978ae}%
  \BibitemOpen
  \bibfield  {author} {\bibinfo {author} {\bibfnamefont {S.~R.}\ \bibnamefont
  {Coleman}},\ }\bibfield  {title} {\emph {\bibinfo {title} {{The Uses of
  Instantons}},}\ }\bibfield  {booktitle} {\emph {\bibinfo {booktitle}
  {{Instantons in gauge theories}}},\ }\href@noop {} {\bibfield  {journal}
  {\bibinfo  {journal} {Subnucl. Ser.}\ }\textbf {\bibinfo {volume} {15}},\
  \bibinfo {pages} {805} (\bibinfo {year} {1979})}\ \BibitemShut {NoStop}%
\bibitem [{\citenamefont {Caldeira}\ and\ \citenamefont
  {Leggett}(1983)}]{Caldeira1982ann}%
  \BibitemOpen
  \bibfield  {author} {\bibinfo {author} {\bibfnamefont {A.}~\bibnamefont
  {Caldeira}}\ and\ \bibinfo {author} {\bibfnamefont {A.}~\bibnamefont
  {Leggett}},\ }\bibfield  {title} {\emph {\bibinfo {title} {Quantum tunnelling
  in a dissipative system},}\ }\href {\doibase
  https://doi.org/10.1016/0003-4916(83)90202-6} {\bibfield  {journal} {\bibinfo
   {journal} {Ann. Phys.}\ }\textbf {\bibinfo {volume} {149}},\ \bibinfo
  {pages} {374 } (\bibinfo {year} {1983})}\ \BibitemShut {NoStop}%
\bibitem [{\citenamefont {Grabert}\ and\ \citenamefont
  {Weiss}(1984)}]{Grabert1984}%
  \BibitemOpen
  \bibfield  {author} {\bibinfo {author} {\bibfnamefont {H.}~\bibnamefont
  {Grabert}}\ and\ \bibinfo {author} {\bibfnamefont {U.}~\bibnamefont
  {Weiss}},\ }\bibfield  {title} {\emph {\bibinfo {title} {Thermal enhancement
  of the quantum decay rate in a dissipative system},}\ }\href {\doibase
  10.1007/BF01469699} {\bibfield  {journal} {\bibinfo  {journal} {Zeitschrift
  f{\"u}r Physik B Condensed Matter}\ }\textbf {\bibinfo {volume} {56}},\
  \bibinfo {pages} {171} (\bibinfo {year} {1984})}\ \BibitemShut {NoStop}%
\bibitem [{\citenamefont {Grabert}\ \emph {et~al.}(1987)\citenamefont
  {Grabert}, \citenamefont {Olschowski},\ and\ \citenamefont
  {Weiss}}]{Grabert1987meta}%
  \BibitemOpen
  \bibfield  {author} {\bibinfo {author} {\bibfnamefont {H.}~\bibnamefont
  {Grabert}}, \bibinfo {author} {\bibfnamefont {P.}~\bibnamefont {Olschowski}},
  \ and\ \bibinfo {author} {\bibfnamefont {U.}~\bibnamefont {Weiss}},\
  }\bibfield  {title} {\emph {\bibinfo {title} {Quantum decay rates for
  dissipative systems at finite temperatures},}\ }\href {\doibase
  10.1103/PhysRevB.36.1931} {\bibfield  {journal} {\bibinfo  {journal} {Phys.
  Rev. B}\ }\textbf {\bibinfo {volume} {36}},\ \bibinfo {pages} {1931}
  (\bibinfo {year} {1987})}\ \BibitemShut {NoStop}%
\bibitem [{\citenamefont {Grabert}\ \emph {et~al.}(1984)\citenamefont
  {Grabert}, \citenamefont {Weiss},\ and\ \citenamefont
  {H\"anggi}}]{Grabert19842}%
  \BibitemOpen
  \bibfield  {author} {\bibinfo {author} {\bibfnamefont {H.}~\bibnamefont
  {Grabert}}, \bibinfo {author} {\bibfnamefont {U.}~\bibnamefont {Weiss}}, \
  and\ \bibinfo {author} {\bibfnamefont {P.}~\bibnamefont {H\"anggi}},\
  }\bibfield  {title} {\emph {\bibinfo {title} {Quantum Tunneling in
  Dissipative Systems at Finite Temperatures},}\ }\href {\doibase
  10.1103/PhysRevLett.52.2193} {\bibfield  {journal} {\bibinfo  {journal}
  {Phys. Rev. Lett.}\ }\textbf {\bibinfo {volume} {52}},\ \bibinfo {pages}
  {2193} (\bibinfo {year} {1984})}\ \BibitemShut {NoStop}%
\bibitem [{\citenamefont {Freidkin}\ \emph {et~al.}(1986)\citenamefont
  {Freidkin}, \citenamefont {Riseborough},\ and\ \citenamefont
  {H\"anggi}}]{Freidkin1986}%
  \BibitemOpen
  \bibfield  {author} {\bibinfo {author} {\bibfnamefont {E.}~\bibnamefont
  {Freidkin}}, \bibinfo {author} {\bibfnamefont {P.}~\bibnamefont
  {Riseborough}}, \ and\ \bibinfo {author} {\bibfnamefont {P.}~\bibnamefont
  {H\"anggi}},\ }\bibfield  {title} {\emph {\bibinfo {title} {Decay of a
  metastable state: A variational approach},}\ }\href {\doibase
  10.1103/PhysRevB.34.1952} {\bibfield  {journal} {\bibinfo  {journal} {Phys.
  Rev. B}\ }\textbf {\bibinfo {volume} {34}},\ \bibinfo {pages} {1952}
  (\bibinfo {year} {1986})}\ \BibitemShut {NoStop}%
\bibitem [{\citenamefont {Riseborough}\ \emph {et~al.}(1985)\citenamefont
  {Riseborough}, \citenamefont {H\"anggi},\ and\ \citenamefont
  {Freidkin}}]{Risborough1985}%
  \BibitemOpen
  \bibfield  {author} {\bibinfo {author} {\bibfnamefont {P.~S.}\ \bibnamefont
  {Riseborough}}, \bibinfo {author} {\bibfnamefont {P.}~\bibnamefont
  {H\"anggi}}, \ and\ \bibinfo {author} {\bibfnamefont {E.}~\bibnamefont
  {Freidkin}},\ }\bibfield  {title} {\emph {\bibinfo {title} {Quantum tunneling
  in dissipative media: Intermediate-coupling-strength results},}\ }\href
  {\doibase 10.1103/PhysRevA.32.489} {\bibfield  {journal} {\bibinfo  {journal}
  {Phys. Rev. A}\ }\textbf {\bibinfo {volume} {32}},\ \bibinfo {pages} {489}
  (\bibinfo {year} {1985})}\ \BibitemShut {NoStop}%
\bibitem [{\citenamefont {Leggett}(1984)}]{Leggett1984prb}%
  \BibitemOpen
  \bibfield  {author} {\bibinfo {author} {\bibfnamefont {A.~J.}\ \bibnamefont
  {Leggett}},\ }\bibfield  {title} {\emph {\bibinfo {title} {Quantum tunneling
  in the presence of an arbitrary linear dissipation mechanism},}\ }\href
  {\doibase 10.1103/PhysRevB.30.1208} {\bibfield  {journal} {\bibinfo
  {journal} {Phys. Rev. B}\ }\textbf {\bibinfo {volume} {30}},\ \bibinfo
  {pages} {1208} (\bibinfo {year} {1984})}\ \BibitemShut {NoStop}%
\bibitem [{\citenamefont {Ankerhold}\ and\ \citenamefont
  {Pollak}(2007)}]{Ankerhold2007}%
  \BibitemOpen
  \bibfield  {author} {\bibinfo {author} {\bibfnamefont {J.}~\bibnamefont
  {Ankerhold}}\ and\ \bibinfo {author} {\bibfnamefont {E.}~\bibnamefont
  {Pollak}},\ }\bibfield  {title} {\emph {\bibinfo {title} {Dissipation can
  enhance quantum effects},}\ }\href {\doibase 10.1103/PhysRevE.75.041103}
  {\bibfield  {journal} {\bibinfo  {journal} {Phys. Rev. E}\ }\textbf {\bibinfo
  {volume} {75}},\ \bibinfo {pages} {041103} (\bibinfo {year} {2007})}\
  \BibitemShut {NoStop}%
\bibitem [{\citenamefont {Cuccoli}\ \emph {et~al.}(2001)\citenamefont
  {Cuccoli}, \citenamefont {Fubini}, \citenamefont {Tognetti},\ and\
  \citenamefont {Vaia}}]{Cuccoli2001}%
  \BibitemOpen
  \bibfield  {author} {\bibinfo {author} {\bibfnamefont {A.}~\bibnamefont
  {Cuccoli}}, \bibinfo {author} {\bibfnamefont {A.}~\bibnamefont {Fubini}},
  \bibinfo {author} {\bibfnamefont {V.}~\bibnamefont {Tognetti}}, \ and\
  \bibinfo {author} {\bibfnamefont {R.}~\bibnamefont {Vaia}},\ }\bibfield
  {title} {\emph {\bibinfo {title} {Quantum thermodynamics of systems with
  anomalous dissipative coupling},}\ }\href {\doibase
  10.1103/PhysRevE.64.066124} {\bibfield  {journal} {\bibinfo  {journal} {Phys.
  Rev. E}\ }\textbf {\bibinfo {volume} {64}},\ \bibinfo {pages} {066124}
  (\bibinfo {year} {2001})}\ \BibitemShut {NoStop}%
\bibitem [{\citenamefont {Kohler}\ and\ \citenamefont
  {Sols}(2006)}]{Kohler2006ky}%
  \BibitemOpen
  \bibfield  {author} {\bibinfo {author} {\bibfnamefont {H.}~\bibnamefont
  {Kohler}}\ and\ \bibinfo {author} {\bibfnamefont {F.}~\bibnamefont {Sols}},\
  }\bibfield  {title} {\emph {\bibinfo {title} {{Dissipative quantum oscillator
  with two competing heat baths}},}\ }\href@noop {} {\bibfield  {journal}
  {\bibinfo  {journal} {New Journal of Physics}\ }\textbf {\bibinfo {volume}
  {8}},\ \bibinfo {pages} {149} (\bibinfo {year} {2006})}\ \BibitemShut
  {NoStop}%
\bibitem [{\citenamefont {Cuccoli}\ \emph {et~al.}(2010)\citenamefont
  {Cuccoli}, \citenamefont {Del~Sette},\ and\ \citenamefont
  {Vaia}}]{Cuccoli2010dr}%
  \BibitemOpen
  \bibfield  {author} {\bibinfo {author} {\bibfnamefont {A.}~\bibnamefont
  {Cuccoli}}, \bibinfo {author} {\bibfnamefont {N.}~\bibnamefont {Del~Sette}},
  \ and\ \bibinfo {author} {\bibfnamefont {R.}~\bibnamefont {Vaia}},\
  }\bibfield  {title} {\emph {\bibinfo {title} {{Reentrant enhancement of
  quantum fluctuations for symmetric environmental coupling}},}\ }\href@noop {}
  {\bibfield  {journal} {\bibinfo  {journal} {Phys. Rev. E}\ }\textbf {\bibinfo
  {volume} {81}},\ \bibinfo {pages} {041110} (\bibinfo {year} {2010})}\
  \BibitemShut {NoStop}%
\bibitem [{\citenamefont {Kohler}\ \emph {et~al.}(2013)\citenamefont {Kohler},
  \citenamefont {Hackl},\ and\ \citenamefont {Kehrein}}]{Kohler2013ie}%
  \BibitemOpen
  \bibfield  {author} {\bibinfo {author} {\bibfnamefont {H.}~\bibnamefont
  {Kohler}}, \bibinfo {author} {\bibfnamefont {A.}~\bibnamefont {Hackl}}, \
  and\ \bibinfo {author} {\bibfnamefont {S.}~\bibnamefont {Kehrein}},\
  }\bibfield  {title} {\emph {\bibinfo {title} {{Nonequilibrium dynamics of a
  system with quantum frustration}},}\ }\href@noop {} {\bibfield  {journal}
  {\bibinfo  {journal} {Phys. Rev. B}\ }\textbf {\bibinfo {volume} {88}},\
  \bibinfo {pages} {205122} (\bibinfo {year} {2013})}\ \BibitemShut {NoStop}%
\bibitem [{\citenamefont {Rastelli}(2016)}]{Rastelli2016ge}%
  \BibitemOpen
  \bibfield  {author} {\bibinfo {author} {\bibfnamefont {G.}~\bibnamefont
  {Rastelli}},\ }\bibfield  {title} {\emph {\bibinfo {title}
  {{Dissipation-induced enhancement of quantum fluctuations}},}\ }\href@noop {}
  {\bibfield  {journal} {\bibinfo  {journal} {New Journal of Physics}\ }\textbf
  {\bibinfo {volume} {18}},\ \bibinfo {pages} {053033} (\bibinfo {year}
  {2016})}\ \BibitemShut {NoStop}%
\bibitem [{\citenamefont {Maile}\ \emph {et~al.}(2020)\citenamefont {Maile},
  \citenamefont {Andergassen},\ and\ \citenamefont {Rastelli}}]{Maile2020}%
  \BibitemOpen
  \bibfield  {author} {\bibinfo {author} {\bibfnamefont {D.}~\bibnamefont
  {Maile}}, \bibinfo {author} {\bibfnamefont {S.}~\bibnamefont {Andergassen}},
  \ and\ \bibinfo {author} {\bibfnamefont {G.}~\bibnamefont {Rastelli}},\
  }\bibfield  {title} {\emph {\bibinfo {title} {Effects of a dissipative
  coupling to the momentum of a particle in a double well potential},}\ }\href
  {\doibase 10.1103/PhysRevResearch.2.013226} {\bibfield  {journal} {\bibinfo
  {journal} {Phys. Rev. Research}\ }\textbf {\bibinfo {volume} {2}},\ \bibinfo
  {pages} {013226} (\bibinfo {year} {2020})}\ \BibitemShut {NoStop}%
\bibitem [{\citenamefont {Weiss}\ \emph {et~al.}(1987)\citenamefont {Weiss},
  \citenamefont {Grabert}, \citenamefont {H\"anggi},\ and\ \citenamefont
  {Riseborough}}]{Grabert1987}%
  \BibitemOpen
  \bibfield  {author} {\bibinfo {author} {\bibfnamefont {U.}~\bibnamefont
  {Weiss}}, \bibinfo {author} {\bibfnamefont {H.}~\bibnamefont {Grabert}},
  \bibinfo {author} {\bibfnamefont {P.}~\bibnamefont {H\"anggi}}, \ and\
  \bibinfo {author} {\bibfnamefont {P.}~\bibnamefont {Riseborough}},\
  }\bibfield  {title} {\emph {\bibinfo {title} {Incoherent tunneling in a
  double well},}\ }\href {\doibase 10.1103/PhysRevB.35.9535} {\bibfield
  {journal} {\bibinfo  {journal} {Phys. Rev. B}\ }\textbf {\bibinfo {volume}
  {35}},\ \bibinfo {pages} {9535} (\bibinfo {year} {1987})}\ \BibitemShut
  {NoStop}%
\bibitem [{\citenamefont {Langer}(1967)}]{Langer1967}%
  \BibitemOpen
  \bibfield  {author} {\bibinfo {author} {\bibfnamefont {J.}~\bibnamefont
  {Langer}},\ }\bibfield  {title} {\emph {\bibinfo {title} {Theory of the
  condensation point},}\ }\href {\doibase
  https://doi.org/10.1016/0003-4916(67)90200-X} {\bibfield  {journal} {\bibinfo
   {journal} {Annals of Physics}\ }\textbf {\bibinfo {volume} {41}},\ \bibinfo
  {pages} {108 } (\bibinfo {year} {1967})}\ \BibitemShut {NoStop}%
\bibitem [{\citenamefont {Kleinert}(1995)}]{Kleinert1995}%
  \BibitemOpen
  \bibfield  {author} {\bibinfo {author} {\bibfnamefont {H.}~\bibnamefont
  {Kleinert}},\ }\href@noop {} {\emph {\bibinfo {title} {Path Integral in
  Quantum Mechanics, Statistics and Polymer Physics}}},\ \bibinfo {edition}
  {2nd}\ ed.\ (\bibinfo  {publisher} {World Scientific Publishing},\ \bibinfo
  {address} {Singapur},\ \bibinfo {year} {1995})\BibitemShut {NoStop}%
\bibitem [{\citenamefont {Weiss}(2012)}]{Weiss2012}%
  \BibitemOpen
  \bibfield  {author} {\bibinfo {author} {\bibfnamefont {U.}~\bibnamefont
  {Weiss}},\ }\href@noop {} {\emph {\bibinfo {title} {Quantum Dissipative
  Systems}}},\ \bibinfo {edition} {4th}\ ed.\ (\bibinfo  {publisher} {World
  Scientific Publishing},\ \bibinfo {address} {Singapur},\ \bibinfo {year}
  {2012})\BibitemShut {NoStop}%
\bibitem [{Note1()}]{Note1}%
  \BibitemOpen
  \bibinfo {note} {The full Hamiltonian is displayed in Appendix \ref
  {App:A}}\BibitemShut {NoStop}%
\bibitem [{Note2()}]{Note2}%
  \BibitemOpen
  \bibinfo {note} {We find the bounce path by inserting the ansatz $
  x_{cl}(\tau )=\protect \frac {1}{\beta }\DOTSB \sum@ \slimits@ _{l} x_{l}
  e^{i\omega _{l}\tau } / \beta $ into the action $S$ and minimizing it with
  respect to $x_l$.}\BibitemShut {Stop}%
\bibitem [{Note3()}]{Note3}%
  \BibitemOpen
  \bibinfo {note} {In the limit $\Sigma \ll V_0$, the action is closely related
  to the problem of the double well studied in \cite {Maile2020} because the
  potentials are equal in the region $x<x_m$.}\BibitemShut {Stop}%
\bibitem [{\citenamefont {Weiss}\ \emph {et~al.}(1984)\citenamefont {Weiss},
  \citenamefont {Riseborough}, \citenamefont {H\"anggi},\ and\ \citenamefont
  {Grabert}}]{Weiss1984}%
  \BibitemOpen
  \bibfield  {author} {\bibinfo {author} {\bibfnamefont {U.}~\bibnamefont
  {Weiss}}, \bibinfo {author} {\bibfnamefont {P.}~\bibnamefont {Riseborough}},
  \bibinfo {author} {\bibfnamefont {P.}~\bibnamefont {H\"anggi}}, \ and\
  \bibinfo {author} {\bibfnamefont {H.}~\bibnamefont {Grabert}},\ }\bibfield
  {title} {\emph {\bibinfo {title} {Energy loss in quantum tunnelling},}\
  }\href {\doibase https://doi.org/10.1016/0375-9601(84)90577-2} {\bibfield
  {journal} {\bibinfo  {journal} {Physics Letters A}\ }\textbf {\bibinfo
  {volume} {104}},\ \bibinfo {pages} {10 } (\bibinfo {year} {1984})}\
  \BibitemShut {NoStop}%
\bibitem [{\citenamefont {Maleeva}\ \emph {et~al.}(2018)\citenamefont
  {Maleeva}, \citenamefont {Gr{\"u}nhaupt}, \citenamefont {Klein},
  \citenamefont {Levy-Bertrand}, \citenamefont {Dupree}, \citenamefont {Calvo},
  \citenamefont {Valenti}, \citenamefont {Winkel}, \citenamefont {Friedrich},
  \citenamefont {Wernsdorfer}, \citenamefont {Ustinov}, \citenamefont
  {Rotzinger}, \citenamefont {Monfardini}, \citenamefont {Fistul},\ and\
  \citenamefont {Pop}}]{Pop2018}%
  \BibitemOpen
  \bibfield  {author} {\bibinfo {author} {\bibfnamefont {N.}~\bibnamefont
  {Maleeva}}, \bibinfo {author} {\bibfnamefont {L.}~\bibnamefont
  {Gr{\"u}nhaupt}}, \bibinfo {author} {\bibfnamefont {T.}~\bibnamefont
  {Klein}}, \bibinfo {author} {\bibfnamefont {F.}~\bibnamefont
  {Levy-Bertrand}}, \bibinfo {author} {\bibfnamefont {O.}~\bibnamefont
  {Dupree}}, \bibinfo {author} {\bibfnamefont {M.}~\bibnamefont {Calvo}},
  \bibinfo {author} {\bibfnamefont {F.}~\bibnamefont {Valenti}}, \bibinfo
  {author} {\bibfnamefont {P.}~\bibnamefont {Winkel}}, \bibinfo {author}
  {\bibfnamefont {F.}~\bibnamefont {Friedrich}}, \bibinfo {author}
  {\bibfnamefont {W.}~\bibnamefont {Wernsdorfer}}, \bibinfo {author}
  {\bibfnamefont {A.}~\bibnamefont {Ustinov}}, \bibinfo {author} {\bibfnamefont
  {H.}~\bibnamefont {Rotzinger}}, \bibinfo {author} {\bibfnamefont
  {A.}~\bibnamefont {Monfardini}}, \bibinfo {author} {\bibfnamefont
  {M.}~\bibnamefont {Fistul}}, \ and\ \bibinfo {author} {\bibfnamefont
  {I.}~\bibnamefont {Pop}},\ }\bibfield  {title} {\emph {\bibinfo {title}
  {Circuit quantum electrodynamics of granular aluminum resonators},}\ }\href
  {\doibase 10.1038/s41467-018-06386-9} {\bibfield  {journal} {\bibinfo
  {journal} {Nat. Commun.}\ }\textbf {\bibinfo {volume} {9}},\ \bibinfo {pages}
  {3889} (\bibinfo {year} {2018})}\ \BibitemShut {NoStop}%
\bibitem [{\citenamefont {Nguyen}\ \emph {et~al.}(2019)\citenamefont {Nguyen},
  \citenamefont {Lin}, \citenamefont {Somoroff}, \citenamefont {Mencia},
  \citenamefont {Grabon},\ and\ \citenamefont {Manucharyan}}]{Manucharyan2019}%
  \BibitemOpen
  \bibfield  {author} {\bibinfo {author} {\bibfnamefont {L.~B.}\ \bibnamefont
  {Nguyen}}, \bibinfo {author} {\bibfnamefont {Y.-H.}\ \bibnamefont {Lin}},
  \bibinfo {author} {\bibfnamefont {A.}~\bibnamefont {Somoroff}}, \bibinfo
  {author} {\bibfnamefont {R.}~\bibnamefont {Mencia}}, \bibinfo {author}
  {\bibfnamefont {N.}~\bibnamefont {Grabon}}, \ and\ \bibinfo {author}
  {\bibfnamefont {V.~E.}\ \bibnamefont {Manucharyan}},\ }\bibfield  {title}
  {\emph {\bibinfo {title} {High-Coherence Fluxonium Qubit},}\ }\href {\doibase
  10.1103/PhysRevX.9.041041} {\bibfield  {journal} {\bibinfo  {journal} {Phys.
  Rev. X}\ }\textbf {\bibinfo {volume} {9}},\ \bibinfo {pages} {041041}
  (\bibinfo {year} {2019})}\ \BibitemShut {NoStop}%
\bibitem [{\citenamefont {Rastelli}\ \emph {et~al.}(2015)\citenamefont
  {Rastelli}, \citenamefont {Vanevi{\'c}},\ and\ \citenamefont
  {Belzig}}]{Rastelli2015}%
  \BibitemOpen
  \bibfield  {author} {\bibinfo {author} {\bibfnamefont {G.}~\bibnamefont
  {Rastelli}}, \bibinfo {author} {\bibfnamefont {M.}~\bibnamefont
  {Vanevi{\'c}}}, \ and\ \bibinfo {author} {\bibfnamefont {W.}~\bibnamefont
  {Belzig}},\ }\bibfield  {title} {\emph {\bibinfo {title} {Coherent dynamics
  in long fluxonium qubits},}\ }\href {\doibase
  https://doi.org/10.1088/1367-2630/17/5/053026} {\bibfield  {journal}
  {\bibinfo  {journal} {New J. Phys.}\ }\textbf {\bibinfo {volume} {17}},\
  \bibinfo {pages} {053026} (\bibinfo {year} {2015})}\ \BibitemShut {NoStop}%
\bibitem [{\citenamefont {Maile}\ \emph {et~al.}(2018)\citenamefont {Maile},
  \citenamefont {Andergassen}, \citenamefont {Belzig},\ and\ \citenamefont
  {Rastelli}}]{Maile2018}%
  \BibitemOpen
  \bibfield  {author} {\bibinfo {author} {\bibfnamefont {D.}~\bibnamefont
  {Maile}}, \bibinfo {author} {\bibfnamefont {S.}~\bibnamefont {Andergassen}},
  \bibinfo {author} {\bibfnamefont {W.}~\bibnamefont {Belzig}}, \ and\ \bibinfo
  {author} {\bibfnamefont {G.}~\bibnamefont {Rastelli}},\ }\bibfield  {title}
  {\emph {\bibinfo {title} {Quantum phase transition with dissipative
  frustration},}\ }\href {\doibase 10.1103/PhysRevB.97.155427} {\bibfield
  {journal} {\bibinfo  {journal} {Phys. Rev. B}\ }\textbf {\bibinfo {volume}
  {97}},\ \bibinfo {pages} {155427} (\bibinfo {year} {2018})}\ \BibitemShut
  {NoStop}%
\bibitem [{\citenamefont {Kechedzhi}\ and\ \citenamefont
  {Smelyanskiy}(2016)}]{Smelyanskiy2016}%
  \BibitemOpen
  \bibfield  {author} {\bibinfo {author} {\bibfnamefont {K.}~\bibnamefont
  {Kechedzhi}}\ and\ \bibinfo {author} {\bibfnamefont {V.~N.}\ \bibnamefont
  {Smelyanskiy}},\ }\bibfield  {title} {\emph {\bibinfo {title} {Open-System
  Quantum Annealing in Mean-Field Models with Exponential Degeneracy},}\ }\href
  {\doibase 10.1103/PhysRevX.6.021028} {\bibfield  {journal} {\bibinfo
  {journal} {Phys. Rev. X}\ }\textbf {\bibinfo {volume} {6}},\ \bibinfo {pages}
  {021028} (\bibinfo {year} {2016})}\ \BibitemShut {NoStop}%
\bibitem [{Note4()}]{Note4}%
  \BibitemOpen
  \bibinfo {note} {Note that for $\Sigma \rightarrow 0$ the steepest decent
  approximation is no longer valid, because the negative eigenvalue $\lambda
  _1^{(B)}$ approaches zero. However, for the minimal asymmetry parameter
  $\Sigma =0.01 V_0$, we use in this paper, the negative eigenvalue is
  $|\lambda _1^{(B)}/(m\omega _0^2)|\approx 0.01$ and the approximation still
  justified}\BibitemShut {NoStop}%
\end{thebibliography}%

\end{document}